\documentclass[12pt,preprint]{aastex}

\usepackage{amssymb}
\usepackage{lscape}
\usepackage{longtable}

\def\lsim{\mathrel{\rlap{\lower4pt\hbox{\hskip1pt$\sim$}}
    \raise1pt\hbox{$<$}}}                
\def\gsim{\mathrel{\rlap{\lower4pt\hbox{\hskip1pt$\sim$}}
    \raise1pt\hbox{$>$}}}                

\shorttitle{The Stellar Mass Density at $z\approx 5$}
\begin{document}


\title{A New Measurement of the Stellar Mass Density at z$\simeq$5: Implications
for the Sources of Cosmic Reionization}


\author {D.~P.~Stark \altaffilmark{1}, 
A.~J.~Bunker \altaffilmark{2}, 
R.~S.~Ellis \altaffilmark{1},
L.~P.~Eyles \altaffilmark{2},
M.~Lacy \altaffilmark{3}
}

\altaffiltext{1}{Department of Astronomy, California Institute of Technology,
MS 105-24, Pasadena, CA 91125; dps@astro.caltech.edu}
\altaffiltext{2}{School of Physics, University of Exeter, 
Stocker Road, Exeter EX4\,4QL, U.K.}
\altaffiltext{3}{Spitzer Science Center, California Institute of Technology,
MC-220-6, 1200 E. California Blvd, Pasadena, CA 91125, U.S.A.}

\begin{abstract} 
We present a new measurement of the integrated stellar mass per comoving 
volume at redshift 5 determined via spectral energy fitting drawn from a sample 
of 214 photometrically-selected galaxies with $z'_{850LP}< $26.5 in the southern 
GOODS field.  Following procedures introduced by \cite{Ey05}, 
we estimate stellar masses for various sub-samples for which reliable and 
unconfused Spitzer IRAC detections are available. A spectroscopic sample 
of 14 of the most luminous sources with  $\overline{z}=4.92$ provides a firm 
lower limit to the stellar mass density of 1 $\times\rm~10^6\,M_{\odot}$ 
Mpc$^{-3}$. Several galaxies in 
this sub-sample have masses of order 10$^{11}\,M_{\odot}$ implying significant
earlier activity occurred in massive systems. We then consider a larger sample 
whose photometric redshifts in the publicly-available GOODS-MUSIC catalog 
lie in the range 4.4 $<z<$ 5.6. Before adopting the GOODS-MUSIC photometric
  redshifts, we check the accuracy of their photometry and explore the 
possibility of contamination by low-z galaxies and low-mass stars.  
After excising probable stellar contaminants and 
using the  $z'_{850LP}-J$ color to exclude any remaining foreground red galaxies, we conclude 
that 196 sources are likely to be at $z\simeq$ 5. The implied mass 
density from 
the unconfused IRAC fraction of this sample, scaled to the total available, 
is $\rm~6\times10^6\,M_{\odot}$ Mpc$^{-3}$.  We discuss the uncertainties 
as well as the likelihood that we have underestimated the true mass density. 
Including fainter and quiescent sources the total integrated density could be 
as high as $\rm~1\times10^7\,M_{\odot}$ Mpc$^{-3}$. Even accounting for 25\% 
cosmic variance within a single GOODS field, such a high mass density only 
1.2 Gyr after the Big Bang has interesting consequences for the implied 
past average star formation during the period
when cosmic reionization is now thought to have taken place. Using the 
currently 
available (but highly uncertain) rate of decline in the star formation
history over 5 $<z<$ 10, a better fit is obtained for the assembled 
mass at $z\simeq$ 5 if we admit significant dust extinction at early
times or extend the luminosity function to very faint limits.
An interesting consequence of the latter possibility is an abundant 
population of low luminosity sources just beyond the detection limits 
of current surveys. As mass density estimates improve at $z\simeq$ 5-6, 
our method is likely to provide one of the tightest constraints on the 
question of whether star forming sources were responsible for reionizing 
the Universe. 

\end{abstract} 

\section{Introduction}
\label{sec:intro}

Finding the sources responsible for cosmic reionization is now the active
frontier in studies of galaxy formation. A number of independent arguments
are focusing efforts on searches for star forming galaxies in the redshift 
interval 5 $<z<$ 10. Studies of the optical depth in Lyman $\alpha$ absorption 
probed by high resolution spectra of the most distant quasars suggest an 
upward transition in the neutral fraction beyond $z\simeq$ 5.5 \citep{Fan06}; 
these data suggest reionization was just ending at $z\simeq$ 6. In 
contrast, the optical depth of microwave photons to electron 
scattering derived from 
the angular power spectrum of the WMAP polarization-temperature 
cross-correlation function \citep{Sp06} places a valuable upper 
bound on the reionization process corresponding to $z\simeq$ 10-20. 

Over the past several years, the quest to observe the most distant
galaxies in the Universe has rapidly expanded to the point where the 
discovery of $z\simeq5-6$ star-forming galaxies has now become routine. 
Deep imaging surveys with the {\em Hubble Space Telescope (HST)} and 
8-10 meter ground based telescopes have uncovered hundreds of galaxies 
at $z\simeq 5$ \citep{Iwata,Bremer} and $z\simeq 6$ 
\citep{Bu04,Di05,Bou06} via the Lyman 
break galaxy (LBG) technique pioneered by Steidel and collaborators 
to identify star-forming galaxies at $z\approx 3-4$ \citep{Ste96,Ste99}.   

The consensus emerging from these studies, however, is that abundance of 
{\it luminous} galaxies is substantially {\it less} at $z\approx 6$ 
than at $z\approx 3$ \citep{St03,Bu04,Di05,Bou06}.  If this trend  
continues to fainter systems and higher redshift, then it may prove 
challenging to explain 
the earlier star formation activity necessary to fulfill reionization in 
the redshift interval 5 $<z<$ 10 implied by the quasar and WMAP studies
\citep{Bu04}.  However, it has been suggested that the evolution
in the galaxy luminosity function between $z = 3$ and $z = 6$ is luminosity 
dependent: although the entire luminosity function is not yet 
well-constrained at $z\approx 6$, intrinsically fainter galaxies 
appear to become more abundant at earlier times \citep{Bou06}.  If 
this is the case, then the bulk of reionizing photons could come from 
lower luminosity galaxies not yet adequately probed in deep surveys.
 
As the redshift boundary of cosmic reionization narrows, so it becomes
crucial to improve our understanding of the cosmic star formation
history in the corresponding time interval. Unfortunately however, 
confirming even the most luminous sources in the range 7 $<z<$ 10 is 
challenging for current facilities. Although some candidate $z\simeq 7-10$ 
galaxies have been identified in ACS and lensed surveys (Bouwens et al
2004b, Bouwens et al 2005, Richard et al 2006, Stark et al. 2007), these 
are generally too faint for spectroscopic study. The situation may not
significantly improve for several years.

This paper explores a more practical approach for constraining the
amount of star formation prior to $z\simeq$ 5-6, namely the measurement
of the integrated stellar mass density at this epoch. Following the 
idea originally discussed by \cite{SE2006}, the stellar mass density 
at $z\simeq$ 5-6 must represents the integral of past activity. With 
adequate precision, such estimates can be used to independently
verify the claimed decline in overall star formation to
$z\simeq$ 10 and to assess whether the past activity is sufficient
for cosmic reionization.

The approach is made practical by the remarkable progress recently
made in estimating stellar masses at high redshift via the use of
the Infrared Array Camera (IRAC, \citealt{Fazio04}) onboard the
{\em Spitzer Space Telescope}. \cite{Eg04} first demonstrated
the technique for one of the most distant known sources: a multiply-imaged 
pair with a photometric redshift of $z\simeq$ 6.8. \cite{Ey05}
later extended the technique for two spectroscopically-confirmed galaxies
at $z$ = 5.8, demonstrating the presence of massive galaxies 
($M_{\rm stellar}>10^{10}\,M_{\odot}$) with evolved stellar populations 
of ages $\gg 100$\,Myr. 

The IRAC filters at $3.6-8.0\,\mu$m probe the rest-frame optical 
at $z\approx 5-6$, providing a valuable indicator of established
stellar populations and, indirectly, hinting at vigorous star formation 
activity at $z> 6$. Combining these data with deep broadband 
optical photometry from {\em HST} and 8-10 meter class ground based 
telescopes, spectral energy distributions (SEDs) can be compared with 
population synthesis models to constrain the age, star formation history and
stellar masses of galaxies. The initial discovery of massive 
($10^{10} M_{\odot}$) galaxies at $z\simeq 6$ presented in 
\cite{Ey05} was subsequently confirmed 
by the independent analysis of \cite{Y05}. More recently, \cite{M05} 
identified a galaxy in the Hubble Ultra Deep Field (UDF) with a photometric 
redshift of $z\simeq 6.5$ (but see also the recent paper by \citealt{Dun06}). 
If this high redshift is correct, then the MIPS 
and IRAC detections imply a very massive system of 
$M_{\rm stellar}>10^{11}\,M_{\odot}$, providing further evidence for 
significant star formation activity at $z>6$ \citep{P05}. 

The studies of galaxy masses published thus far have focused on only
a few individual systems. Although some studies (e.g. Stark \& Ellis 2006)
have attempted to infer the contribution of past star formation to cosmic
reionization, without knowing how typical such massive galaxies are, it is 
difficult to make precise statements. Clearly what is needed is a {\it census}
of the assembled stellar mass at high redshift. A comoving {\it stellar mass
density} can be directly compared with various models of earlier star 
formation.

In a companion paper, we compute the stellar mass density at 
$z\approx 6$ from the $i'$-band dropouts in GOODS-South \citep{Ey06}. 
A similar study of $i'$-drops was conducted in \cite{Y06}.
However, the surface density of $i'$-band dropout galaxies at $z\approx 6$ 
with {\em Spitzer} detections is low. A more 
statistically-meaningful sample can be found using the $z\approx 5$ 
$v$-band dropouts. The age of the Universe at this time is only marginally
older (1.2 Gyr c.f. 0.95 Gyr) yet larger, more representative, samples
are available.  In this paper we will examine the stellar mass density at
$z\approx $5 using sources to a limiting magnitude of $z'_{850LP}\approx 26.5$
selected from the Great Observatories Origins Deep Survey (GOODS, 
\citealt{G04a}). We present an analysis of various 
subsamples at $z\approx 5$ drawn from a total of $\simeq$214 $v$-band
dropouts. The goal of the study is to establish whether the assembled stellar 
mass at $z\simeq 5$ is consistent with current (and admittedly uncertain) 
estimates of the preceding star formation activity. If not, this might be taken
to imply a significant component of star formation is missing, 
occurring either at lower intrinsic luminosities, obscured by dust, 
or at uncharted epochs ($z>10$). 

A plan of the paper follows. In $\S$\ref{sec:GOODSdataset}, 
$\S$\ref{sec:photom}, and $\S$\ref{sec:selection}, we introduce 
the various imaging and spectroscopic 
datasets, the photometric procedures and the selection of various subsamples 
of $z\simeq$ 5 galaxies. We describe the derivation of the stellar masses and
comment on the uncertainties in $\S$\ref{sec:mass}. In $\S$\ref{sec:sfrdens}, 
we examine the implications for the star formation history at earlier times.

We adopt a cosmology consistent with the initial WMAP data release
 \citep{Sp03}: a $\Lambda$-dominated, flat universe with 
$\Omega_{\Lambda}=0.7$, $\Omega_{M}=0.3$ and $H_{0}=70\,h_{70} 
{\rm km\,s}^{-1}\,{\rm Mpc}^{-1}$. All magnitudes in this paper are 
quoted in the AB system \citep{Oke83}.

\section{The GOODS-S Dataset\label{sec:GOODSdataset}}

In this paper, we continue our analyses of the 
Great Observatories Origins Deep Survey (GOODS). GOODS
aims to bring together the most powerful space and
ground-based facilities to study the high-redshift universe 
across a wide range of wavelengths. We focus on the southern
GOODS field which has the greatest amount of multi-wavelength data
essential for reliable stellar masses. The GOODS-S survey area
covers a total of 160 arcmin$^2$ and is centered on the {\em Chandra} Deep 
Field South (CDF-S; \citealt{giacconi}).

\subsection{{\em ACS} Imaging}
\label{subsec:ACS}

Deep optical imaging of GOODS-S has been obtained with the 
Advanced Camera for Surveys ({\em ACS}, \citealt{ford03}) instrument onboard
 {\em HST} as part of a Treasury Program \citep{G04}.  
The Wide Field Camera on ACS has a field of 
202$\times$202 arcsec$^2$ and a pixel scale of 0\farcs 05.   
The GOODS-South field was observed in the F435W ($B$-band), 
F606W ($v$-band), F775W (SDSS-$i'$) and F850LP (SDSS-$z'$) 
broad-band filters for 3, 2.5, 2.5 and 5 orbits, respectively
over 16 pointings.

Here we present an analysis of $z\simeq 5$ galaxies making use
of the publicly-available version-1.0 data-release of the {\em ACS} GOODS
data\footnote{available from {\tt
ftp://archive.stsci.edu/pub/hlsp/goods/}}.
The reduced data have been `drizzled' onto a large grid 
made up of 18 sections with a pixel scale of 0$\farcs$03. 
Each section comprises an image of 8192 $\times$ 8192 pixels in size.  

\subsection{Ground-Based Near-infrared Imaging}
\label{subsec:nearIR}

Deep near-infrared observations of most of the GOODS-S field 
were obtained with the ISAAC camera on the Very Large Telescope 
(VLT) at the ESO Paranal Observatory as part of the ESO Large Programme: 
LP168.A-0485(A) (PI: C. Cesarsky).  The publically available 
version-1.5 data release includes 24 fully reduced ISAAC/VLT 
pointings in the $J$ and $K_{s}$-bands\footnote{available from
{\tt http://www.eso.org/science/goods/releases/20050930/}}, covering 
$\approx 160\,{\rm arcmin}^2$.
Additional details of the observations are to be presented
in Vandame et al.\ ({\em 2006, in prep}). The VLT images have a pixel
scale of 0\farcs15, a factor of five times larger than the drizzled 
ACS pixels. The median exposure times are 11.3 ksec in $J$,
and  17.9 ksec in $K_s$. 

\subsection{Spectroscopy}
\label{subsubsec:spectra}

We also use publicly-available spectroscopy from the GOODS
team to identify confirmed $z\approx 5$ galaxies for futher study.
Multi-object spectroscopy was performed on the GOODS-S field with the
FORS2 instrument mounted at the Kueyen Unit Telescope of the VLT at
ESO's Cerro Paranal Observatory as part of the ESO/GOODS Large Program
LP170.A-0788 (PI Cesarsky).  Details of the survey are presented in
\cite{Vanzella02,Vanzella05}.  The primary selection criteria for
placing objects on the slitmask was ($i'_{775W}-z'_{850LP})>0.6$ and
$z'_{850LP} < 25.0$; objects with $0.45 < (i'_{775W}-z'_{850LP}) < 0.6$ were
placed on the slitmask with lower priority.  We make use of the
VLT/FORS2 spectroscopic catalogs from the version-2.0
release which provide 725 unique redshift assignments with quality
flags A, B, or C (where A=solid redshift, B=likely redshift, C=potential
redshift).
 
\subsection{{\em Spitzer} Imaging}
\label{subsubsec:spitzer}

{\em Spitzer} images of GOODS-S were obtained with the Infrared 
Array Camera (IRAC) and Multiband Imaging Photometry for Spitzer (MIPS)
cameras on the {\em Spitzer} Space Telescope as part of the ``Super Deep'' 
Legacy programme (PID 169, Dickinson et al.\ {\em in prep}, Chary et al., 
{\em in prep}).  The IRAC camera comprises four channels, each with 
a $256^2$ InSb array imaging a $5.2'\times 5.2'$ field with a pixel 
size of $\approx 1\farcs22$. Images were taken through four broad-band infrared
filters, with central wavelengths at approximately $\lambda_{\rm
  cent}=3.6\,\mu$m, $4.5\,\mu$m, $5.6\,\mu$m and $8.0\,\mu$m (channels
1--4), and widths of $\Delta\lambda_{\rm  FWHM}=0.68,0.87,1.25,2.53\,\mu$m 
respectively.  The total exposure time in each channel is
$\approx$\,86\,ksec, depending on location. The data were taken in two epochs,
with the telescope roll angle differing by 180$^{\circ}$. 
In the first epoch, each filter covered a $10.0'\times 10.0'$ area in GOODS-S; 
however, the area covered by channels 1 and 3 ($3.6\,\mu$m and $5.6\,\mu$m) 
was offset by 6.7 arcminutes from that covered by channels 2 and 4 
($4.5\,\mu$m and $8.0\,\mu$m). Hence, only a portion
of the GOODS-S field was observed in all 4 filters after the 
first epoch of observations.  In the second epoch, the area 
covered by channels 1 and 3 in the first epoch was observed with 
channels 2 and 4 and vice versa. A central overlap region appeared
in both epochs, and this deeper area intentionally contains the Hubble Ultra
Deep Field (HUDF, \citealt{Beck_udf,Bu04}).

We analyze the publicly available {\em Spitzer} mosaics 
from the first and second epochs of the observations of GOODS-S 
\footnote{available from {\tt http://data.spitzer.caltech.edu/popular/goods}}.
The data reduction pipeline employs a 
`multidrizzle' technique similar to that used successfully on 
{\em HST}/ACS GOODS data. This provides combined images with a pixel 
scale of 0\farcs6. The magnitudes listed in this paper are 
determined from this `drizzled' data.  We use the updated ``Super Deep'' 
epoch 1 images from the third data release (DR3) and the Super Deep 
epoch 2 images from the second data release (DR2).

\section{Photometric Samples}
\label{sec:photom}

The photometry we compute in this section will be used for two independent 
samples of $z\simeq 5$ objects: a small sample of spectroscopically confirmed 
galaxies and a larger sample of photometrically selected galaxies.  The spectroscopic
sample will provide a robust lower limit to the $z\simeq 5$ stellar 
mass density whereas the photometric sample will provide a more 
representative estimate of the integrated mass density.  To obtain 
stellar masses of individual galaxies, we must have accurate photometry 
for both samples as well as photometric redshifts for the photometric sample.  
The reliability of the photometric redshifts is especially 
crucial since contamination by low-redshift interlopers could 
seriously skew our estimates of the total mass.

We obtain photometric redshifts from the GOODS MUSIC photometric catalog 
of GOODS-S \citep{graz06}.  This catalog uses uses 13-band SEDs from {\em HST}
/ACS and {\em Spitzer}/IRAC photometry along with ground-based $U$, $J$, 
\& $K_S$ to derive photometric redshifts. Before adopting the GOODS MUSIC photometric 
redshifts, we verify the accuracy of the photometry in the GOODS MUSIC catalog 
(discussed below) and test the reliability of their photometric redshifts which 
we discuss in $\S4.1$.   

ACS photometry was obtained from the GOODS team r1.1 catalog
\footnote{available from {\tt http://archive.stsci.edu/prepds/goods}}.
The photometric zeropoints 
adopted in the catalog on the AB magnitude system are 25.653, 26.493, 
25.641, and 24.843 for the $B_{435W}$-band, $v_{606W}$-band, 
$i'_{775W}$ band, and $z'_{850LP}$-band, respectively. 
We have corrected for the small amount of foreground Galactic extinction
using the {\it COBE}/DIRBE \& {\it IRAS}/ISSA dust maps
of \cite{sfd98}; for the GOODS-S field, 
selective extinction is given by E($B-V$) = 0.008 mag. Magnitudes are 
measured in 0\farcs50-diameter apertures.  Total magnitudes are derived
from the aperture magnitudes by correcting for the small amount of light falling 
outside the aperture: 0.14, 0.15, and 0.20  magin the $v_{606W}$,$i'_{775W}$, 
and $z'_{850LP}$-bands, respectively \citep{sir05}.   We note that 
GOODS website implies that the SExtractor parameter PHOT\_APERTURES measures
the {\it radius} of the photometric aperture, when it in fact measures the 
{\it diameter}.  The correct interpretation has been applied to our dataset.  

Near-infrared photometry was performed with $1''$-diameter apertures
using the ground-based near-infrared ISAAC images.  The center 
of the photometric aperture was taken from the centroid of the 
GOODSr1.1 catalog.  The seeing varied
across the ISAAC field as different tiles were taken over many
nights, so we determined separate aperture corrections from unresolved
sources for each tile. For the $J$- and $K_s$-band images the seeing
is typically good (${\rm FWHM}=0\farcs4-0\farcs5$), and the aperture
corrections are $\approx 0.3-0.5$\,mag, determined from bright but
unsaturated isolated stars measured in $6''$-diameter apertures.
The 3$\sigma$ limiting AB-magnitudes in a $1''$-diameter aperture 
are $J\approx 26.4$ and $K_{S}\approx 25.7$, although 
these vary over the field because of different exposure times and 
seeing conditions.

The details of the photometric analysis of the Spitzer images used in
this paper are nearly identical to those presented in \cite{Ey05}.  In
order to maximize the signal-to-noise ratio ($S/N$) and minimize
possible confusion with other foreground objects, we used a
photometric aperture of diameter $\approx 1.5\times {\rm FWHM}$ for
the IRAC images, appropriate for unresolved objects (our compact
souces are essentially unresolved at IRAC resolution, see e.g.\ 
\citealt{Bremer}). 
The aperture diameters were 4, 4, 5 \& 6 `drizzled'
pixels for the 4 channels (3.6, 4.5, 5.6 \& 8.0\,$\mu$m),
corresponding to $2\farcs4$, $2\farcs 4$, $3\farcs 0$, \& $3\farcs 7$.
We used the IRAF {\tt digiphot.phot} package to measure the enclosed
flux at the coordinates determined by the ACS GOODS-r1.1 catalogs,
taking the residual background from an annulus between $12''$ and
$24''$ radius. We applied aperture corrections to compensate for the
flux falling outside the aperture: these were $\approx 0.7$\,mag for
the IRAC data, as determined from bright but unsaturated point sources
in the images using large apertures.

The noise for each of the four channels was checked in two different
ways.  First, we derived an estimate based on a Poisson model using
the detector gain, number of frames combined, and the background
counts (adding back the zodiacal background estimate subtracted by the
pipeline but recorded in the header). Secondly, we measured the
standard deviation in background counts of the images.  As the
mosaicking process introduces correlations between pixels, we also made
noise estimates using the individual pipeline basic calibrated data
(BCD) images and assuming it decreased as the square root of the
number of frames. These estimates lead to $3\sigma$ limiting AB
magnitudes of 26.5 and 26.1 using $2\farcs 4$-diameter apertures in
channels 1 and 2, respectively, and 23.8 and 23.5 in $3\farcs
0$ and $3\farcs7$-diameter apertures in channels 3 and 4, 
respectively.  There will be additional background fluctuations 
caused by faint galaxies (i.e. confusion noise), which will increase 
the noise. Both methods produce consistent estimates.

The low spatial resolution of {\em Spitzer} results in frequent 
blending between nearby sources, making accurate photometry
of individual objects difficult.  We took great effort
to ensure that objects in our sample were not contaminated 
by neighboring bright foreground sources.  We approach the 
IRAC contamination in slightly different ways for the 
different subsamples of z$ \simeq$ 5 objects.  Details
are provided in $\S$\ref{sec:selection}. 

We find that our photometry is consistent with that in the GOODS 
MUSIC cataog.  The standard deviation between our 
photometry and the GOODS MUSIC photometry is 0.13 mags, 0.03 mags, 
0.13 mags for the z', J, and K band. This increases to 0.36 mags for the 
3.6 micron IRAC photometry. 

\section{Selection of z$\simeq$5 Galaxies}
\label{sec:selection}

\subsection{The Photometric Sample}

We make use of the extensive database of photometric
redshifts in the publicly-available GOODS-MUSIC catalog 
\citep{graz06} to construct a sample of $z\simeq$ 5 candidates.  
Details of the procedure used to compute the photometric redshifts 
are discussed in \cite{graz06}. As described below, we 
based our photometric selection on the GOODS-MUSIC catalog 
rather than upon a more traditional $v$-band dropout technique 
\citep{Bremer,G04,Y05} on account of the improved performance in various
tests.  The principle difference is that the former method is based on 
fitting the entire SED.  

First, we consider the fidelity of the GOODS-MUSIC selection
of $z\simeq$ 5 galaxies with respect to the VLT spectroscopic results
of \cite{Vanzella02, Vanzella05}. 21 galaxies within our $z\simeq 5$ 
spectroscopic sample (see next section) have photometric 
redshifts in the GOODS-MUSIC catalog. 18 of these ($>$85\%)
have photometric redshifts in the 4.4 $<z<$ 5.6 range 
with an average absolute scatter of $<|z_{spec}-z_{phot}|>$=0.07.
Two of the three objects for which the photometric redshifts 
fail completely (e.g photometric redshifts of $z\simeq 1-2$)
have spectroscopic redshift quality grades of C: here it is 
possible that the photometric redshifts are actually correct.  
This test suggests the SED-fitting process is reasonably accurate.
 
\begin{figure}
\plotone{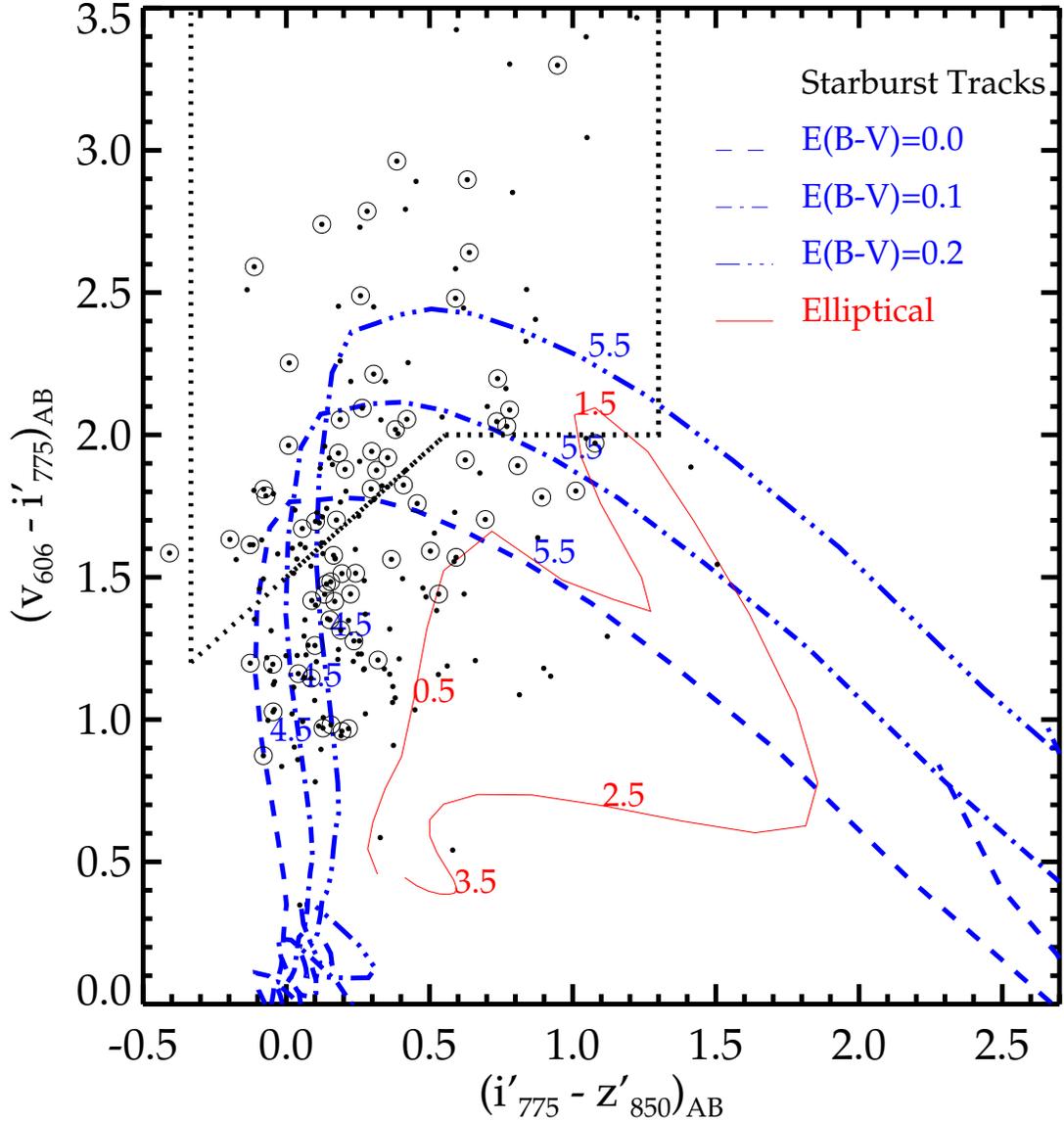}
\caption{$(v_{606W} - i'_{775W})$ vs. $(i'_{775W} - z'_{850LP})$ 
colors of z$\simeq$5 candidates in GOODS-S.  We construct a sample
of 214 objects with photometric redshifts between
4.4 $<$ z $<$ 5.6 from the GOODS-MUSIC catalog (solid black circles).  
After removing stellar contaminants, low-z 
interlopers, and objects blended in Spitzer images, 72 objects 
with $z'_{850LP} < 26.5$ remain; these objects are marked 
with an additional circle.  Although many objects in the sample 
fall just outside of the $v$-band dropout selection window used 
by \cite{G04} to select $z\simeq 5$ galaxies (demarcated by dotted line), 
redshift  tracks generated from starburst templates from \cite{bc03} 
illustrate that their rest-frame UV colors are consistent with the 
z$\simeq$5 interpretation.  These tracks assume an age of 100 Myr and 
a constant star formation rate with E(B$-$V)=0.0,0.1,0.2 (blue dashed
line, dashed-dotted, dashed-triple-dotted, respectively).  Redshifting an 
elliptical galaxy template \citep{cww} to z=0-4 (red solid line), we see that 
old galaxies at z$\simeq$1.5 could contaminate our sample. 
\label{plot:izvi_tracks_music}}
\end{figure}

A further verification of the reliability of the photometric catalog
concerns the implied rest-frame colors. Adopting a magnitude 
limit of $z'_{850LP} < 26.5$ (the 50\% completeness limit for 
unresolved sources in GOODS \citealt{G04}), we find 214 objects 
with photometric redshifts between 4.4 $<z<$ 5.6. Their rest-frame 
UV colors are in uniformly good agreement with those expected 
from the locus of star-forming galaxies  at $z\simeq$ 5 
(Figure \ref{plot:izvi_tracks_music}).

We find that only 42\% of the objects in our photometric catalog would 
have been selected in the traditional \cite{G04} $v$-drop 
method.  Examining the redshift tracks, it is clear that the $v$-drop 
method misses a significant fraction of $z=4.5-5.5$ star-forming 
galaxies (Figure \ref{plot:izvi_tracks_music}).  
This region of color-color space is not included in the 
traditional $v$-drop method to minimize the inclusion of low 
redshift contaminating galaxies.  The GOODS-MUSIC photo-z sample  
(along with the criteria we impose below) represents an improvement to 
the traditional $v$-drop selection criteria as it takes the entire SED into
consideration in assessing an object's redshift.

While the GOODS-MUSIC photometric redshifts appear to be an 
excellent predictor of the true redshift, we remain vigilant to the 
possibility of a few catastrophic failures. The point is critical as the presence 
of any residual low redshift or stellar objects that are very bright at 
{\em Spitzer} wavelengths could lead to a significant overestimate of 
the stellar mass density. Recognizing there is a danger of removing
true $z\simeq $5 sources, we conclude it is better to err on the
conservative side.

Because of their red colors, low-mass stars are a common contaminant 
of photometrically-located high redshift galaxy samples. Bright stars can be removed 
from high-z galaxy samples by selecting unresolved objects in 
the HST/ACS images. However, this technique begins to fail  
at fainter magnitudes as extragalactic objects may appear unresolved 
if observed at low S/N.  Alternatively, stellar contaminants 
can be selected from our sample on the basis of their optical through
near-infrared.  We fit the SEDs of all objects in the photometric 
catalog with M, L, and T dwarf stellar templates \citep{Leggett02, West05,
Kraus06}.  We construct 
a list of stellar contaminants by examining each object well-fit 
with stellar colors, only including sources without extended 
emission.  Our final list consists of 11 stars (5\% of the total sample) 
with $z'_{850LP}=24.3 - 26.5$, each of which we remove from our 
photo-z sample.      

Low-redshift galaxies with intrinsically red colors arising from dust 
extinction or an old stellar population commonly contaminate traditional 
dropout samples because their (v$_{606W}$-$i'_{775W}$) 
colors are similar to those of  $z\simeq$ 5 star-forming objects.
By considering the shape of the entire SED, low-z interlopers can often 
be identified and removed from high-redshift dropout samples.
Since the GOODS-MUSIC photometric redshifts are computed using 
the entire SED, we expect the contamination rate from low-z galaxies to 
be low.  Nevertheless, we believe it is important to explore the 
possibility that 
low-redshift galaxies may remain in the GOODS-MUSIC sample and examine 
the effects that possible contaminants may have on our final results.

A simple way to estimate the contamination rate from low-z galaxies is 
to measure the rest-frame UV$-$optical colors of each of the objects in our 
sample.  Unextincted star-forming objects at $z\simeq$ 5 typically have
spectra that are roughly flat in f$_\nu$ (as a function of wavelength) 
between the Lyman break and rest-frame $\simeq$4000~\AA.  In 
contrast, the colors of low-redshift contaminants are red in all filters.  
To quantify the expected difference in  rest-frame UV$-$optical colors 
between $z\simeq$ 5 sources and possible low-z contaminants, we examined 
a set of \cite{bc03} population synthesis models.  Elliptical galaxies 
at $z\simeq $1-2 with ages $>$2 Gyr have ($z'_{850LP}-J)$ colors that 
vary between
1.4 - 1.6; whereas young ($\simeq$ 100 Myr) star-forming
galaxies at $z\simeq $5 with E(B$-V$)=0.0-0.2 have (z$'_{850LP}-J)$ colors 
ranging between $-$0.1 and $-$0.3  Accordingly, to test the  
low-z contamination rate, we adopt a (z$'_{850LP}-J) >$ 1.0 threshold 
\footnote{The precise value of this color discriminant is not critical 
in defining the final sample} above which we consider galaxies to be possible 
low-z interlopers.  Seven objects in our photo-z sample satisfy this color criterion.  
Six of the seven objects are relatively faint in the IRAC filters, and thus will hardly
contribute to the total stellar mass of the sample. One of the objects 
($23$\_$18055$), however, is very bright in the near and mid-infrared 
(m$_{3.6\mu m}$=21.1); if at z$\simeq$5, its best-fit stellar mass would 
be 2$ \times$ 10$^{12}$ M$_\odot$. Given that no objects are identified at 
$z> 4$ with 
stellar masses above 3$\times$10$^{11}$ M$_\odot$ in the 0.8 sq deg UKIDSS survey 
\citep{Dun06}, we conclude that it is much more realistic to adopt a low-z 
interpretation for this object. 
So as not to bias our total mass estimates we remove the seven objects with 
(z$'_{850LP}-J$)$ >$ 1.0 from our photo-z sample, leaving 196 objects.

The final photometric sample is that for which the {\em Spitzer} IRAC
images reveal a clear, unconfused, detection. Reliable stellar masses cannot
otherwise be determined. We examined the {\em Spitzer} images
of each of the 196 $z\simeq $5 candidates, classifying them as
either (1) isolated and detected, (2) undetected, (3) confused or (4)
hopelessly confused.  In the subsequent analysis, we consider only those 
objects that are detected and isolated.  Of the 196 candidates,
72 are sufficiently uncontaminated to allow reliable estimates of the 
stellar mass. 

Table~\ref{tab:vdrop_photom} lists the measured optical through infrared AB 
magnitudes (corrected to approximate total magnitudes through an aperture 
correction), colors, and photometric redshifts for the remaining 72 $z\simeq 5$ objects.

\subsection{The Spectroscopic Sample}

\begin{figure}
\plotone{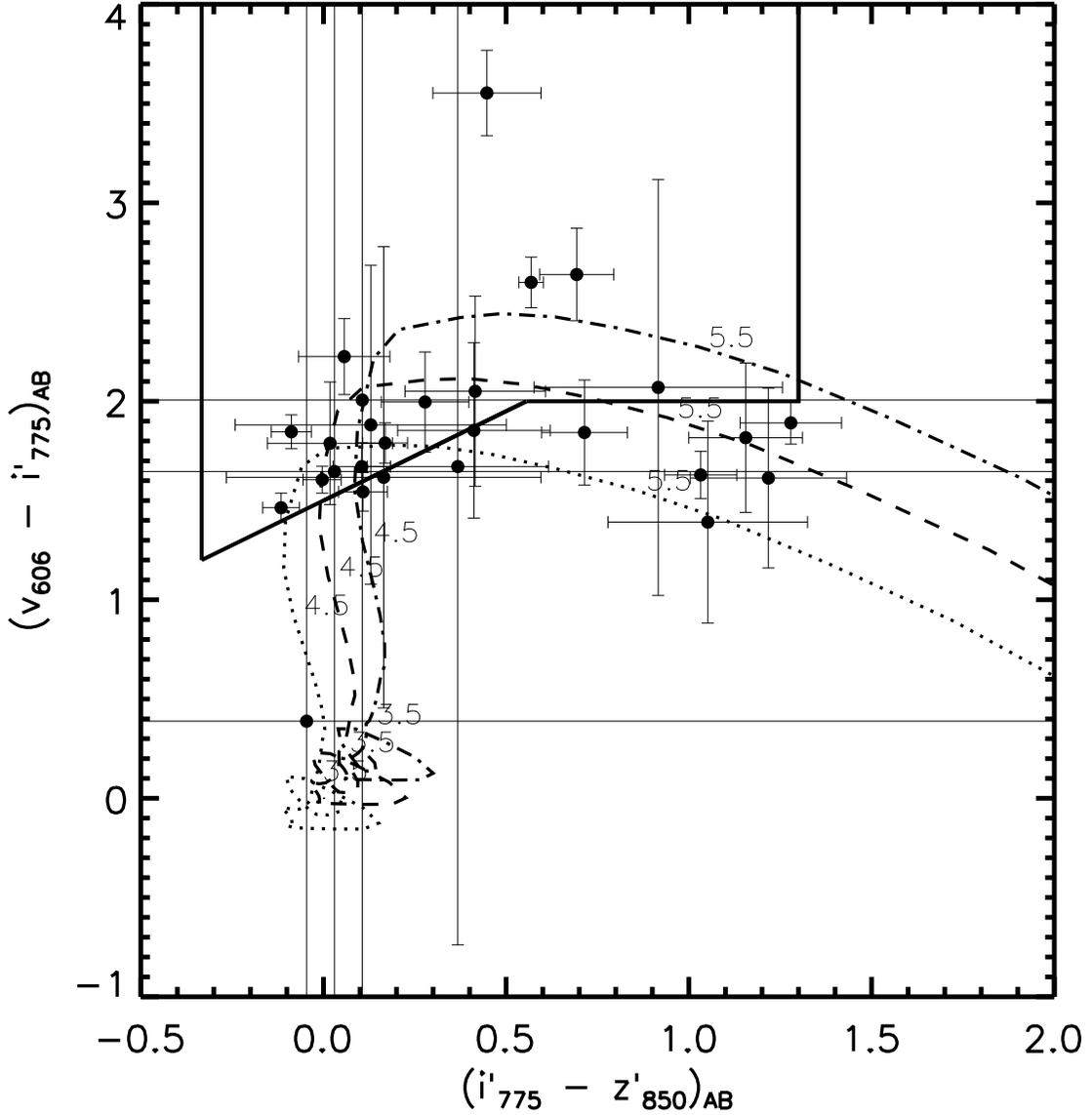}
\caption{($v_{606W} - i'_{775W}$) vs. ($i'_{775W} - z'_{850LP}$) 
colors of 30 galaxies with FORS2/VLT
spectroscopic redshifts of 4.4 $<$ z $<$ 5.6.  The 
$v$-drop selection window described in \cite{G04} 
is overlaid with a solid line.  Starburst 
redshift tracks are identical to those 
described in Figure \ref{plot:izvi_tracks_music}. Removing
objects that either do not satisfy the magnitude limit of $z'_{850LP}<$26.5 
or do not have the rest-frame UV colors expected for a $z\simeq 5$ LBG 
leaves a sample of 25 objects.  
}
\label{plot:tracks_spec}
\end{figure}

The FORS2/VLT spectroscopic survey of the GOODS-S field identified 30
unique galaxies in the $4.4 < z < 5.6$ redshift range.  The quality
flags associated with the redshift assignments range from A (solid) to
C (potential).  As with the photometric sample, we adopt a magnitude limit of 
$z'_{850LP}<26.5$; this requirement excises one object (35\_11820) from the sample. 
Given the possibility of uncertainties in the spectroscopic
identification of those sources with C-grade redshifts, we examined their
rest-frame ultraviolet colors ($v_{606W} - i'_{775W}$) vs. ($i'_{775W} - 
z'_{850LP}$)  as an additional criterion for selection (Figure~\ref{plot:tracks_spec}).
     
Of the 29 remaining FORS2 galaxies with spectroscopic redshifts of $z\simeq 5$,
only 17 would be selected as $v$-drops using the \cite{G04} selection
criteria. An additional 8 of the spectroscopically-confirmed
$z\simeq5$ galaxies fall very near the $v$-drop selection
window in the ($v_{606W} - i'_{775W}$) vs.\ ($i'_{775W} - z'_{850LP}$) 
color-color plot. As their colors are
consistent with the Bruzual \& Charlot redshift tracks (plotted in
Figure~\ref{plot:tracks_spec}) we include them in this sample.
Three objects are apparently undetected in the {\em ACS} images of GOODS-S. 
Without the availability of rest-frame UV colors, we cannot confirm
that the objects are truly located $z\simeq $5 via the presence of
the Lyman break; we presume these were serendipitous detections
and exclude them from the final spectroscopic sample.  The final object 
(22\_15184) is formally a B-dropout; its relatively blue ($v_{606W} - i'_{775W}$) 
color is inconsistent with that expected from a v-drop.  At the object's purported 
redshift ($z=5.08$) Ly$\alpha$ falls in the $i'$-band, making the intrinsic
($v_{606W} - i'_{775W}$) {\it bluer} than what is measured.  Given the peculiar
colors, we exclude it from the spectroscopic catalog.

As before, we examined the {\em Spitzer} images of each of the 25
spectroscopically-confirmed galaxies for detections and the degree
of confusion. These classifications are shown in Table~\ref{tab:spec_sample}.
Five objects were isolated and detected in the Spitzer images, four sources 
were hopelessly confused, and the remaining 16 objects were marginally confused. 
 For the 16 partially confused galaxies, we attempted 
to subtract the contribution from contaminating sources using the `GALFIT'
software package \citep{Peng02}; this was deemed worthy given the
need to maximize the information from the limited spectroscopic data.

GALFIT constructs a two-dimensional model of the data according to 
specified input parameters (e.g., magnitude, position, axis ratio, effective radius), 
performs a convolution with the instrument point spread function (PSF), and fits 
the result to the data through an iterative $\chi^2$ minimization process. 
We determined the PSF for each  epoch and channel of the `drizzled' 
{\em Spitzer} images by stacking 4 bright but isolated stars. For each 
galaxy we assumed a generalized S\'ersic surface brightness profile,
where $\log I\propto r^{1/n}$, and fit for the shape and index $n$.

An automated script was developed to run GALFIT three times per source
on a $12\times 12$\,arcsec$^2$ region surrounding the contaminated
object for the IRAC images.  In the first iteration, we held all source parameters 
fixed in the fitting process except the source magnitude, which was estimated
from the SExtractor source detection software version-2.2.1 \citep{Bertin96}
.  All other input source parameters (e.g. position, axis
ratio, position angle, effective radius, S\'ersic parameter) were
estimated from a fit to the VLT $K_s$-band image.  The higher spatial
resolution of the $K_s$-band allows better deblending and more
accurate centroids to be derived for confusing objects in the IRAC
images. In the second GALFIT iteration, we again determined input
parameters using our fit to the $K_s$-band image, but this time we 
allowed all parameters to vary.  In the final iteration, we obtained the 
initial parameters by applying SExtractor to the IRAC channel 1 
($3.6\,\mu$m) image and allowing all parameters to vary.  For each 
source, we selected the most successful of the three GALFIT runs, 
on the basis of visual inspection of the residual image and the 
$\chi^2$ value for the fit. Those sources (7 out of 16) for which GALFIT failed 
to satisfactorily subtract contaminating emission were removed from the 
sample (see Table~\ref{tab:spec_sample}). The photometry of the 
remaining 14 galaxies are described in Table~\ref{tab:spec_photom}. 

\section{Stellar Mass Determination}
\label{sec:mass}

Although we have removed many sources from the original spectroscopic
and photometric samples, it is worth reminding that the degree of confusion
in the IRAC images should, on average, be completely independent of the stellar
mass of the $z\simeq$ 5 galaxy. Confusion in the IRAC images will
normally arise from the overlapping isophotes of unrelated sources. Thus, 
if sources are believed to be at $z\simeq$ 5 on the basis of a spectroscopic 
redshift or the ACS and $K$-band photometric SED, we can rescue a reasonable 
estimate of their contribution to the stellar mass density by scaling that 
determined for the unconfused sample using the relative numbers.

\subsection{Masses for the Spectroscopic Sample}

For those galaxies with confirmed spectroscopic redshifts,
we estimate stellar masses by fitting population synthesis 
models to the observed SEDs.  Applying this technique to galaxies
for which the redshift is unknown may lead to significant uncertainty 
in the derived properties \citep{Bundy05,Sh05}, hence for 
the photometrically-selected sample, we infer stellar mass by applying 
the median mass-to-light ratio derived from the spectroscopic sample. 

We proceed as \cite{Ey05} by fitting the latest \cite{bc03} 
stellar population synthesis models to the observed SEDs.  We use the
Padova evolutionary tracks preferred by \cite{bc03}.  The models
utilise 221 age steps from $10^5$ to $2\times 10^{10}$\,yr,
approximately logarithmically spaced.  For each source, we do not
include age steps in excess of the age of the universe at $z\simeq 5$
($\simeq$\,1.2 Gyr). Models with \cite{salpeter55} initial mass
functions (IMF) were selected; although we also considered the effect of
adopting a \cite{chabrier} IMF. There are 6900 wavelength steps, with
high resolution (FWHM 3\,\AA ) and 1\,\AA\ pixels evenly-spaced over
the wavelength range of 3300\,\AA\ to 9500\,\AA\ (unevenly spaced
outside this range).  From the full range of metallicities offered by
the code, we considered both solar and 0.25 Z$_\odot$ models. 
 From several star formation histories available, a
single stellar population (SSP -- an instantaneous burst), a constant
star formation rate (SFR), and exponentially decaying ($\tau$) SFR
models with e-folding decay timescales
$\tau$=10, 30, 50, 70, 100, 200, 300, 500, 1000 Myr were used. 

For each of the galaxies in our sample, the
filters were corrected to their rest-frame wavelengths by the
appropriate redshift factor.  The measured flux was folded through the
filter transmission profiles, and the best-fit age model was computed
by minimizing the reduced $\chi^{2}$, using the measured errors on the
magnitudes.  The number of degrees of freedom is the number of
independent data points (magnitudes in different wavebands) minus
the number of parameters that we are fitting. The
Bruzual \& Charlot spectra are normalized to an initial mass of
$1\,M_{\odot}$ for the instantaneous burst (SSP) model, and an SFR of
$1\,M_{\odot}\,{\rm yr}^{-1}$ for the continuous star formation model.
The fitting routine returned the normalisation for the model which was
the best-fit to the broad band photometry (i.e., minimized the reduced
$\chi^{2}$).  This normalization was then used to calculate the
corresponding best-fit total mass using the luminosity distance 
for the redshift of each source.  When considering models other than 
an SSP (instantaneous burst), it was necessary to correct the total 
'mass' values output by the fitting routine.  For a constant SFR 
model, each of these masses needed to be multiplied by the 
corresponding best-fit age, since the B\&C template normalization 
has the mass grow by 1 M$_\odot$ yr$^{-1}$.  For the exponential 
decay models, the returned mass values were corrected by dividing 
by (1$-$e$^{-t/\tau}$), accounting for the decay timescale and the 
normalization of the B\&C models (where M $\longrightarrow$ 1 M$_\odot$
as t $\longrightarrow$ $\infty$). The fits to the B\&C models returned 
the 'total mass' which is the sum of the mass currently in stars,
 in stellar remnants, and in gas returned to the interstellar medium 
by evolved stars.  For each best-fit model, we subsequently calculate the 
mass currently in stars for every galaxy, again using information 
from the B\&C population synthesis code; we use this stellar mass in 
all future analysis.

Although some of our data points (particularly from the {\em HST}/ACS
imaging) have $S/N>10$, we set the minimum magnitude error to be
$\Delta({\rm mag}) = 0.1$ to account for calibration uncertainties.
Futhermore, we do not include data with photometric error above 0.72 mags 
(1.5 $\sigma$).  

The presence of a strong spectral line in one of the broadband
filters could significantly skew the SED fitting.  Seven of the 
15 galaxies in our spectroscopic sample show powerful Ly$\alpha$ emission.
Using the FORS2 spectra, we compute and subtract the Ly$\alpha$ contribution
to the broadband flux; corrections range from $0.01 - 0.1 $ mags
for most sources.  H$\alpha$ contamination could also be a significant issue -
\cite{chary05} claim to find an excess due to
H$\alpha$ in a z=6.5 galaxy (in the 4.5$\mu$m band at that redshift). 
The sources in our sample are likely to have H$\alpha$
emission lines as well, which at z$\simeq$5 fall in either the 3.6 $\mu$m 
or 4.5 $\mu$m IRAC filter. Indeed many of the SED fits discussed below 
show an excess at 3.6 $\mu$m.  Without a direct measure of the H$\alpha$ line 
strengths, we cannot robustly remove the line contamination. 
We estimate that, for most sources, H$\alpha$ contributes 
$\simeq $10-20\% of the measured broadband flux by converting 
the inferred rest-frame UV star formation rate 
to an H-alpha luminosity via empirically-derived 
relations from Kennicutt (1998) assuming  $< 
SFR_{H\alpha}/SFR_{UV}>$=1.5-3 
due to dust extinction in agreement with observations, e.g.  
\cite{Erb03}.  To test the effects that 
H$\alpha$ contamination may have on our sources, we re-fit
all the objects in the spectroscopic sample, omitting the 
flux information at 3.6 $\mu$m for objects with $z<5.2$ and 
at 4.5 $\mu$m for objects with $z>5.2$.  We find that this does not
significantly change the total stellar mass found in our 
spectroscopic sample.

The degeneracies associated with the derived 
best-fit parameters from SED fitting are well known
\citep{Sh01,Pap01,Sh05}.  The uncertainties primarily
stem from a poor knowledge of the form of the star 
formation history, since the best fit age, 
dust extinction and star formation rate rely on 
this \citep{Sh05}. In most cases, the data do not put strong constraints
on the form of the star formation history; hence
each fitted parameter typically has a range of 
values that produce acceptable fits. 

The inferred properties also rely 
on knowledge of the stellar initial mass 
function (IMF).   There is little
observational information constraining the IMF at 
high redshift.  The spectrum of the z=2.7 
gravitationally-lensed LBG cB58 appears to be 
inconsistent with IMFs that have steep high-mass 
slopes or are truncated at high stellar masses 
\citep{Pet00}.  Whether this is typical
among LBGs is unclear. \cite{Pap01} studied the 
effects that varying the IMF have on the best-fit 
parameters. Models with IMFs containing steep high-mass 
slopes (e.g. Scalo, Miller-Scalo) have redder 
integrated spectra and hence younger derived ages 
and lower extinction. As with \cite{Ey05}, we find 
good agreement between the properties inferred using 
a Salpeter IMF and a Chabrier IMF: the best-fit ages are 
nearly the same and the stellar masses are typically
30\% lower when the Chabrier IMF
is used.  Here all masses are quoted for the
Salpeter IMF in order to maintain consistency with 
previous estimates of stellar mass and star formation
rates.  

In Figure \ref{plot:spectra}, we display the best-fit SEDs for each 
of the galaxies in our spectroscopic sample. The best-fitting model 
parameters are presented in Table \ref{tab:mod}. The best fitting 
stellar masses of the galaxies range between 
3$\times$10$^8$ M$_\odot$ and 2$\times$10$^{11}$ M$_\odot$. 
Derived ages span three orders of magnitude, from 
1 Myr to 1.1 Gyr, the age of the Universe at $z\simeq$ 5.
Interestingly, three of our sources have stellar masses
in excess of $10^{11}$  M$_\odot$, values approaching the
high stellar mass for the UDF source located by Mobasher et al (2005).
Our results provides support for at least the presence of
such galaxies even if their abundance remains uncertain.
Moreover, some of the less massive sources can only be fit 
with remarkably young ages ($<$20 Myr) reminiscent of
the lensed star forming source located by \cite{Ellis01}.

The total stellar mass of the subsample of spectroscopic 
galaxies is $5 \times$ 10$^{11}\, M_\odot$.  Clearly 
this estimate is an {\it unrealistic} lower limit 
to the total stellar mass since there are nearly twice as many 
objects known to be at $z\simeq 5$ that certainly have 
stellar mass. 

Uncertainties on this mass arise from two main 
sources.  First, the photometric error for each data-point
in the SED translates into an uncertainty in the inferred 
stellar mass.  Second, there is a range
in acceptable masses that result from varying the age, extinction,
and star formation history.  With regard to the latter,
we follow the approach outlined in \cite{Ey05}
where confidence intervals were explored for two sources.  
We present mass-age confidence intervals for two sources
representative of our sample (Figure \ref{plot:confidence}).
Uncertainties in the inferred stellar mass of individual 
objects in Table 4 range between 30\% and 50\%. Objects detected 
at low signal to noise generally have larger uncertainties.  Given 
the range of uncertainties, it seems reasonable to transfer a 
50\% uncertainty to all of our combined masses.

\begin{figure}
\plotone{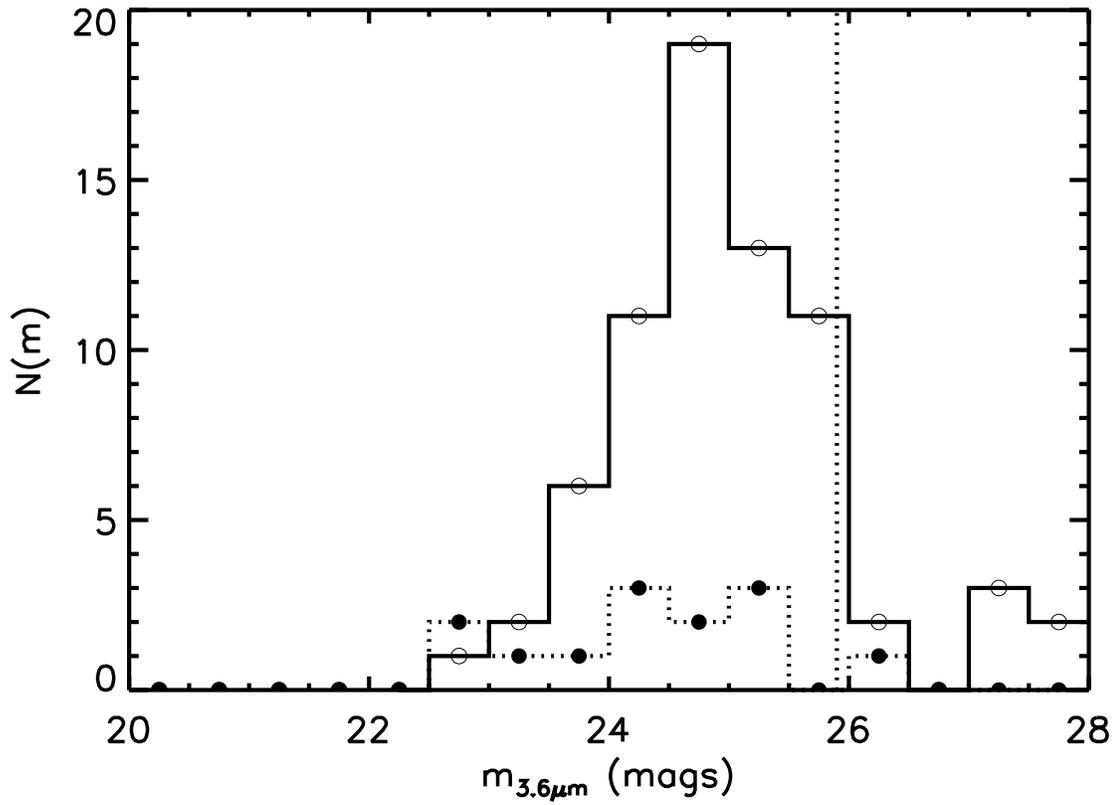}
\caption{Distribution of IRAC 3.6 $\mu$m AB magnitudes for 72 photometrically-
selected z$\simeq$5 candidates (open circles) and 14 
spectroscopically-confirmed z$\simeq$5 galaxies (closed circles).  
The spectroscopic sample contains a larger relative fraction of 
Spitzer bright objects.
}
\label{plot:irac_histo}
\end{figure}

\begin{figure}
\epsscale{.85}
\plotone{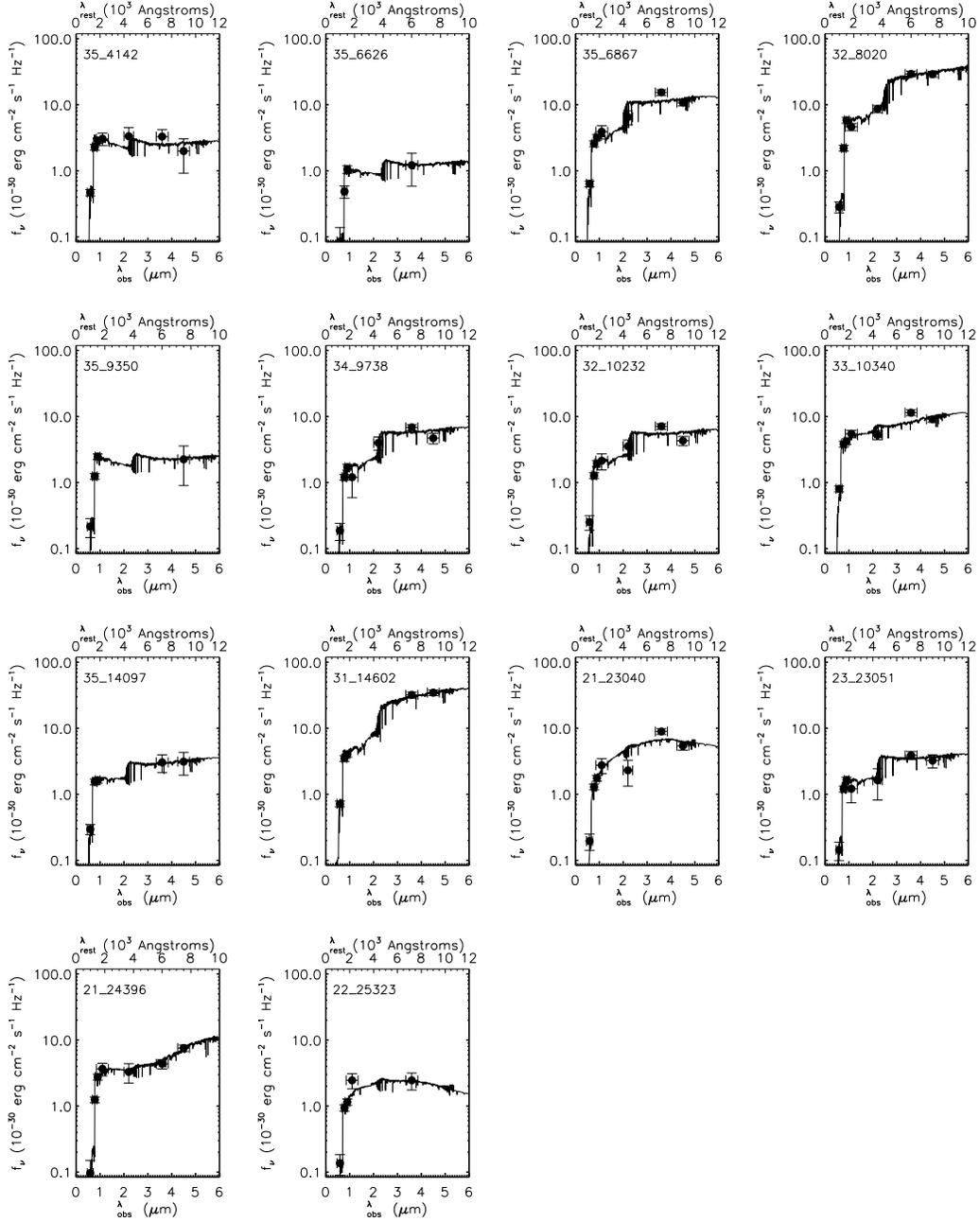}
\caption{Observed and best-fit model \cite{bc03} 
SEDs of 15 spectroscopically-confirmed $z\simeq $5 
galaxies in GOODS-S.  Best-fit model parameters
are presented in Table~\ref{tab:mod}.  Three objects 
(32\_8020, 31\_14602, and 33\_10340) have inferred stellar masses 
above 10$^{11}$ M$_\odot$, and an additional three
objects have inferred stellar masses greater than 10$^{10}$
M$_\odot$.  
}
\label{plot:spectra}
\end{figure}

\begin{figure}
\plottwo{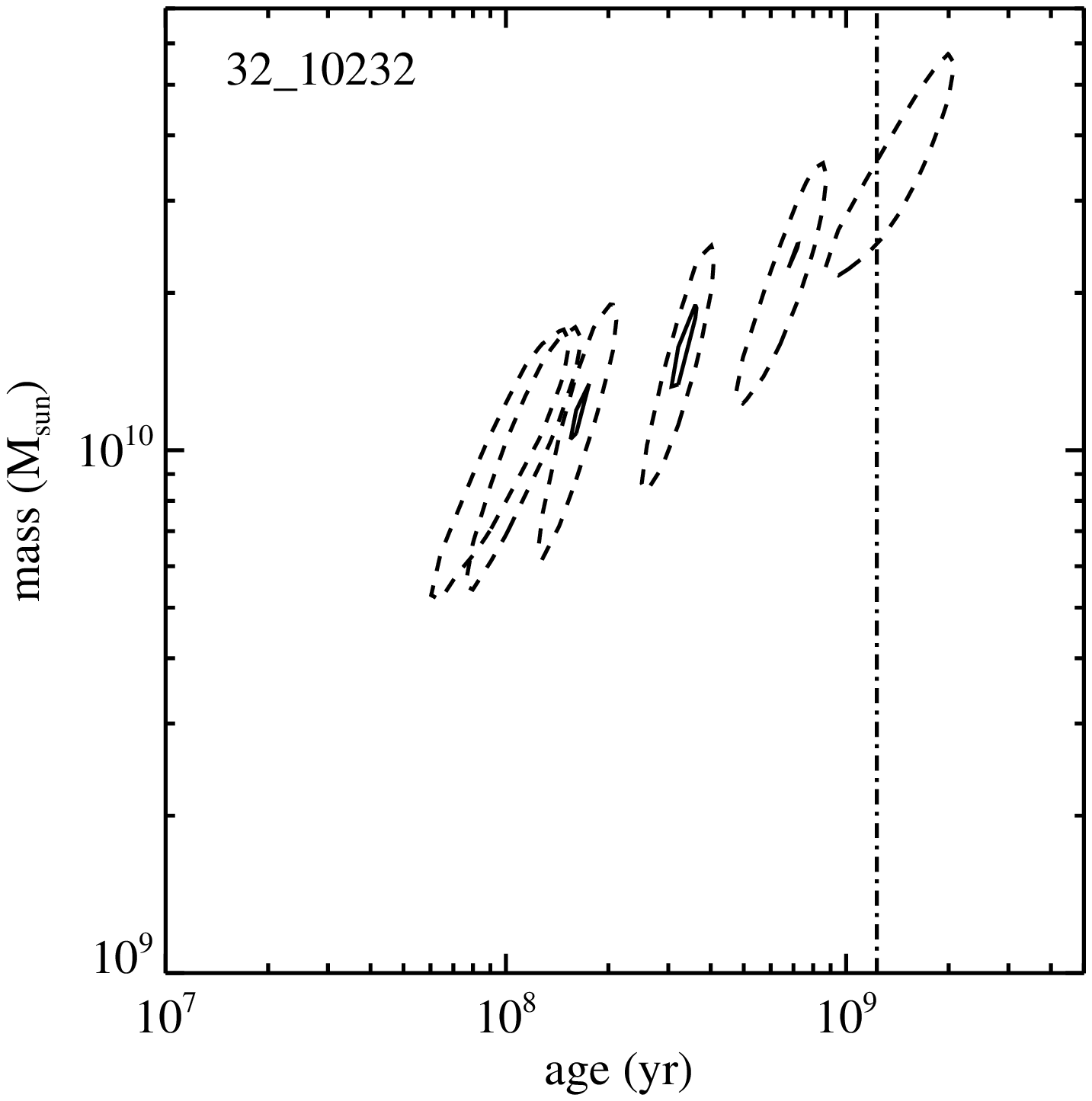}{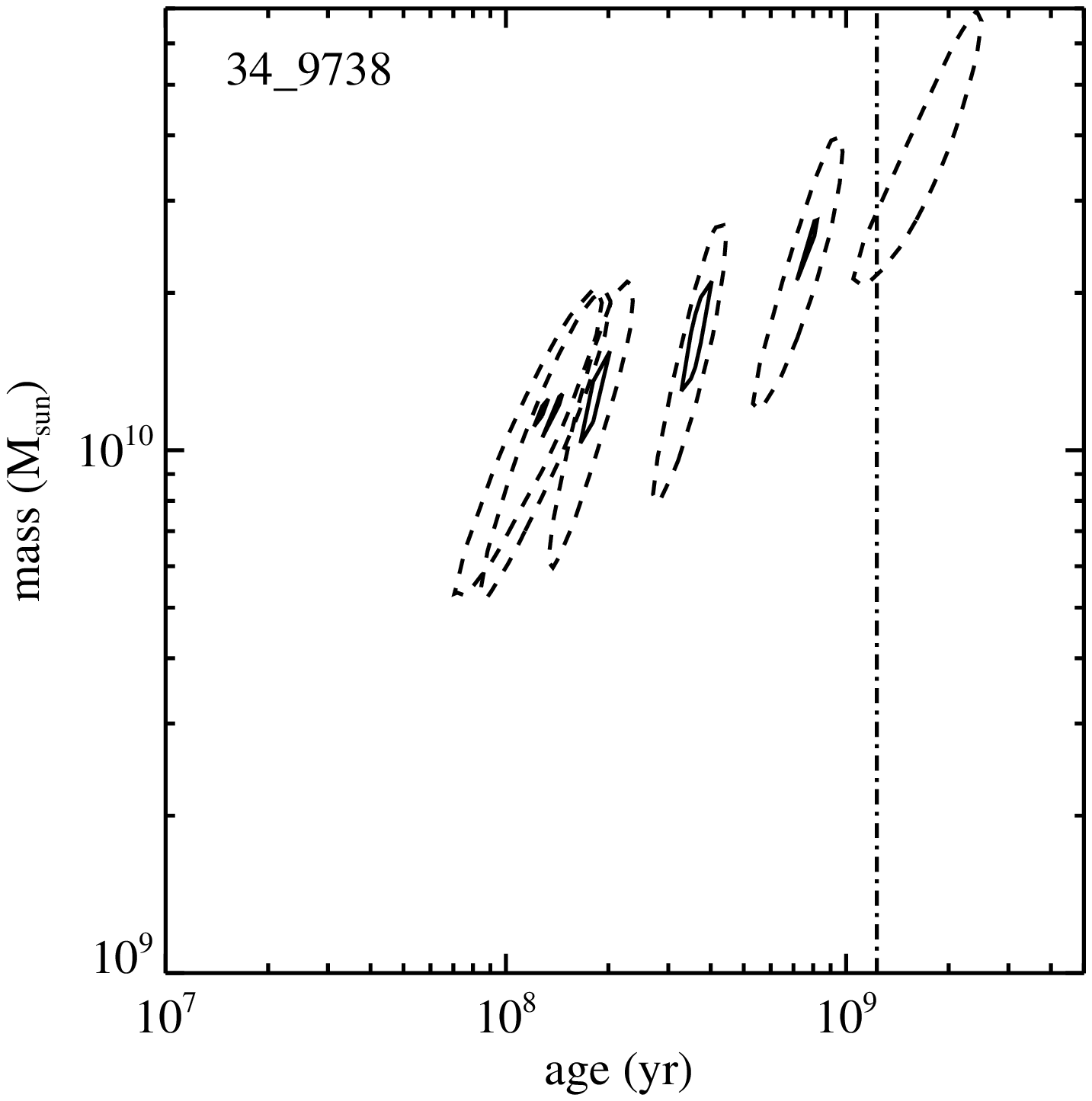}
\caption{Confidence intervals for inferred stellar mass versus age 
for two objects from the spectroscopically-confirmed
z$\simeq$5 objects in GOODS-S.  The ellipses are different assumed 
star formation histories, ranging from an initial burst to continuous 
star formation via a range of exponentially-decaying star formation 
histories. Contours are 68\% confidence (solid line) 
and 95\% confidence (dashed line) corresponding to 
$\Delta$$\chi_{reduced}^2$ = 1, 4 (respectively), 
where $\Delta$$\chi_{reduced}^2$=$\chi_{reduced}^2$ $-$ $\chi_{reduced, min}^2$.  
The vertical dashed-dotted line at the right 
of each plot denotes the age of the universe at the 
source's redshift.  Solutions to the right of this line are 
ruled out.  The typical 68\% confidence 
uncertainties in the stellar mass are are 30-50\%.  
\label{plot:confidence}}
\end{figure}

The FORS2 selection of galaxies was not geared specifically towards
constructing a z$\simeq$5 sample; hence it is important to examine how 
the properties of spectroscopically-confirmed galaxies compares to the 
photometrically-selected sample.  The median rest-frame UV color of the 
spectroscopic sample, ($< i'_{775W}-z'_{850LP}>$)=0.17, is very similar to the 
photometrically-selected sample ($< i'_{775W}-z'_{850LP}>$=0.21).   
The key parameter for determining the stellar mass is the flux in the IRAC filters. 
The histogram of IRAC 3.6 $\mu$m fluxes for the spectroscopic and 
photometric samples is given in Figure \ref{plot:irac_histo}.  The spectroscopic 
sample does contain a larger fraction of Spitzer-bright 
(e.g m$_{3.6\mu m} <$ 23) objects, but this is reasonable if the overall 
rest-frame mass/light distribution is fairly similar across the population. 

\subsection{Masses for the Photometric Sample}

To estimate stellar masses for the photometric sample
we compute the best-fitting rest-frame $V$-band mass
to light ratio of each galaxy that is unconfused in the IRAC 
images and multiply by the luminosity derived from the 
IRAC flux.  

The best-fitting M/L$_V$ is determined for each galaxy from its 
$z'_{850LP}-m_{3.6\mu m}$ color (corresponding to the ratio of 
rest-frame UV and optical fluxes).   If we assume the typical 
galaxy in rest-frame UV selected samples at z$\simeq$5$-$6 has little dust 
as seems reasonable (see Table 4 and Eyles et al. 2006), 
then the $z'_{850LP}-m_{3.6\mu m}$ color is correlated with 
the age of the galaxy and hence, for a given IMF and star formation 
history, its M/L$_V$ ratio.  

This is done for a given galaxy in the sample by first computing 
$z'_{850LP}-m_{3.6\mu m}$ colors for Bruzual-Charlot models 
(redshifted to the galaxy's photometric redshift) with ages ranging 
between 0 and 1.2 Gyr (roughly the lookback time at the redshift of the galaxy).
We then find the model that produces the $z'_{850LP}-m_{3.6\mu m}$ color 
closest to that observed for a given galaxy.  This model is taken to have 
the ``best-fit'' age and M/L$_V$ ratio for this particular galaxy.  At $z\simeq 5$, 
the 3.6$\mu$m IRAC filter covers the rest-frame V-band; hence, we convert the 
3.6 $\mu$m flux to a luminosity (assuming $z=z_{phot}$) and multiply it by the 
best-fitting M/L$_V$ to compute the stellar mass.  For each galaxy, best-fitting 
stellar masses are  computed for the same range of single-component star formation 
histories used to fit the spectroscopic sample.  The stellar mass we assign to each 
galaxy is taken from the star formation history that produces the best-fitting 
$z'_{850LP}-m_{3.6\mu m}$ colors.  We obtain an estimate of the systematic 
uncertainty in the mass by considering the range inferred 
from the different star formation histories and ages that provide a good fit 
(e.g. $\Delta \chi^2$ = $\chi^2 -\chi^2_{min}$ $<$ 1)to the observed $z'_{850LP}-
m_{3.6\mu m}$ color.  

We note that the observed 3.6$\mu$m luminosity is not equivalent to
a rest-frame V-band luminosity for all redshifts.  The 3.6$\mu$m 
band shifts between rest-frame 5500-6700 \AA~ for $z=4.4$-5.5.  To
test the systematic offsets introduced by relying on M/L$_V$ to 
derive masses, we compare the mass of the spectroscopic sample derived 
in the manner described above to the mass from SED fitting.  We 
find the median offset between the two methods is 40\%.  

The total stellar mass extracted from the 72 $z\simeq 5$ sources
that are uncontaminated in the Spitzer images is $5-9\times$10$^{11}$ 
M$_\odot$ with a best-fit value of 7$\times\rm~10^{11}$ M$_\odot$.
The median stellar mass in the sample is 6$\times$10$^9$ M$_\odot$.  
If we make the reasonable assumption that the distribution of stellar masses is 
independent of IRAC contamination, we can estimate the stellar mass in IRAC 
contaminated galaxies by 
multiplying the stellar mass derived from the uncontaminated $z\simeq 5$
galaxies by the ratio of the total number of $z\simeq 5$ sources to the 
number of uncontaminated $z\simeq 5$ sources.  Following this 
reasoning, the total stellar mass for the photometric
sample becomes 2$\times10^{12} M_\odot$.  Taking the full range of single-component 
star formation histories into consideration, this total stellar mass 
could lie between 2 and 3$ \times\rm~10^{12}$ M$_\odot$.   

\subsection{Comoving Mass Densities}

To derive the comoving stellar mass densities from the above totals,
we need to estimate the redshift-dependent selection function in the 
160 arcmin$^2$ GOODS-S field between 4.4 $<$ z$<$ 5.6.
Although the total possible comoving volume is 5.6 $\times$10$^{5}$
Mpc$^{3}$, the effective volume is less than this value due to sample 
incompleteness arising as a result of objects being scattered
faintward of the magnitude limit or out of the color-selection window. 

In order to account for
these luminosity and redshift biases, we compute an effective survey
volume following the approach of \cite{Ste99} using
\begin{equation}
V_{\rm eff}(m)=\int dz\,p(m,z)\,\frac{dV}{dz}
\end{equation}
where $p(m,z)$ is the probability of detecting a galaxy at redshift $z$
and apparent $z'$ magnitude $m$, and $dz\,\frac{dV}{dz}$ is the comoving
volume per unit solid angle in a slice $dz$ at redshift $z=4.4-5.6$. 

We compute the probability function $p(m,z)$ by putting thousands of fake galaxies 
into the GOODS images and recreating a photometric catalog for the 
new image using identical selection parameters used in generating the 
GOODSv1.1 catalogs. The apparent magnitudes of the fake galaxies span 
$z'_{850LP}=22-27$ in steps of $\Delta m$=0.5 and redshifts ranging between 
$z=4$ and $z=6$ in steps of $\Delta z$=0.1.  The sizes of the fake galaxies are 
consistent with distriution of half-light radii derived for $z\simeq 5$ galaxies 
in Ferguson et al. (2004).  The colors of the fake galaxies 
depend on the galaxy redshift and SED.   We adopt the SED of a Bruzual-Charlot 
model with constant star formation history, an age of 100 Myr, and no dust as the 
intrinsic rest-frame SED of the fake galaxies. Allowing for a selective extinction 
of E(B-V)=0.1 in the fake galaxies' SEDs decreases the effective volume by roughly 5\%, 
which would not significantly change our final mass density estimates.  
The colors are then determined for 
galaxies at each redshift in a manner similar to that which we described in $\S5.1$.  
The probability function, $p(m,z)$ is then given by the fraction of fake galaxies 
with apparent magnitude, m, and redshift, z, that are brighter than the magnitude 
limit and satisfy the dropout color selection criteria.  Since our 
selection is based on photometric redshifts, we adopt color-criteria
that are appropriate for our photometric sample ($v_{606W}-i'_{775W}>0.9$ 
and $i'_{775W}-z'_{850LP}>1.3$).  

The effective volume probed is 5.2$\times$10$^5$ Mpc$^3$ at $z'_{850LP}=23$ 
and $z=5$ where we are nearly 100\% complete and falls to 
1.2$\times$10$^5$ Mpc$^3$ at z$'_{850LP}$=26.5. The stellar mass density 
inferred from our $z\simeq$5
candidates is thus $5-8\times$10$^6$ M$_\odot$ Mpc$^{-3}$, with a 
best-fit value of 6$\times$10$^6$ M$_\odot$ Mpc$^{-3}$.  The robust 
lower limit from our spectroscopic sample is 1$\times$10$^6$ M$_\odot$ 
Mpc$^{-3}$.

Our inferred stellar mass density is most likely an underestimate
of the total value at $z\simeq 5$ for several reasons. 
Foremost, the survey is only sensitive to the most luminous and perhaps
most massive galaxies since we only included objects with significance 
above 3$\sigma$ at 3.6$\mu$m. 
  
Second, an additional reservoir of stellar mass may be contained in objects
that are not currently forming stars and hence are very faint in the rest-frame 
ultraviolet.  At $z\simeq 3$, LBGs contribute only 17\% of the stellar
mass density in the most massive sources \citep{vd06}; the 
remaining fraction is likely contained in objects that are not actively forming 
stars.  Although this fraction of quiescent sources is probably much lower at 
earlier times, we conservatively estimate the total mass density could rise
by a further factor of two.

In summary, therefore, we derive a firm lower limit to the stellar
mass density at $z\simeq$ 5 of 1$\times$10$^{6}$ M$_\odot$ Mpc$^{-3}$, 
a reasonable estimate of the total {\it observed} population of 
5-8$\times$10$^{6}$ M$_\odot$ Mpc$^{-3}$ and cannot exclude 
undetected sources which would increase the total to 
1$\times$10$^{7}$ M$_\odot$ Mpc$^{-3}$. Although the overall estimates
span a factor of 2, we emphasize that the spectroscopic sample is
clearly a significant underestimate of the observed population.

\section{Implications for the Previous Star Formation History}
\label{sec:sfrdens}

In the foregoing we have attempted to put the first bounds on the
stellar mass density at $z\simeq$ 5, 1.2 Gyr after the Big Bang
and about 800 Myr after $z\simeq$ 10. We emphasize that there are considerable
uncertainties in the various steps in our analysis. First, to derive stellar mass 
estimates, we had to cull our samples to those with reliable IRAC detections, 
later scaling on the assumption that they represent a fair subset of the
spectroscopic and photometric populations. For our spectroscopic sample,
our fitting procedure gives mass estimates that span a wide range
depending on the assumed star formation history. Finally, we assumed
a median visual mass/light ratio for the photometric sample derived from that for
the spectroscopic sample. 

Probably the dominant error in deriving the total mass density is not the 
numerical scaling factors , but rather 
the intrinsic uncertainty in estimating the masses of individual galaxies. 
Detailed work at lower redshift \citep{Sh05,Bundy05,Pap05} has shown that inferences 
of the stellar mass from 
SED fitting yield mass estimates that contain typical uncertainties of $\simeq $30\%.
The errors certainly increase slightly when considering objects at higher redshift;
however, our error estimates ($\S$5.1) suggest that the stellar mass estimates
of objects in our spectroscopic sample are typically 50\%, possibly more.

Notwithstanding the uncertainties, it is interesting to now consider
the implications of our derived mass density. The star formation rate 
density (SFRD) of bright ($>$ 0.3 L$^\star_{z=3}$)  star-forming galaxies at 
$z\simeq 5-10$ appears to decline continuously toward higher redshift
\citep{Bu04,Bou04b,Bou05}. However, current 
observations may be missing a substantial fraction of star formation either 
because it is enshrouded in dust, too faint to be detected with current 
facilities, or located at redshifts uncharted by current telescopes ($z>10$).  
Comparing the comoving density of assembled stellar mass at $z\simeq 5$ 
with estimates derived from models of the previous star formation history 
enables us to test these possiblities, thereby providing constraints
beyond direct reach of current facilities.

Taking data on the SFRD from the recent literature
\citep{G04,Bou04b,stan04,Bou05,Bou06}, we fit the
redshift dependence with a simple functional form over $z\simeq 5-10$.  
In cases where the SFRD was not evaluated down to the adopted
fiducial luminosity limit (0.1L$^\star$ at z=3), we compute the
additional contribution by integrating the luminosity function assuming 
the Schechter function parameters derived in each paper.  Since the 
data at $z>7$ do not allow for the robust derivation of the form of the
luminosity function, we assume the shape of the luminosity function remains 
constant before $z\simeq 6$.  

\begin{figure}
\plotone{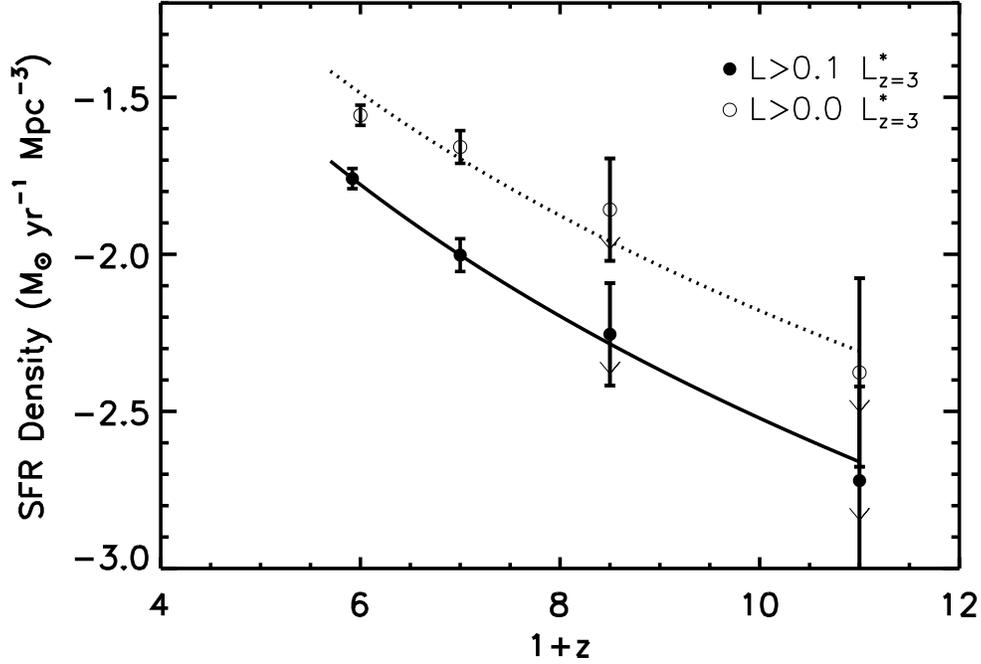}
\caption{Comoving star formation rate density as a function of redshift,
assuming no extinction. The star formation rate densities are derived
from Giavalisco et al. (2004) at $z=5$, Stanway (2005) and Bouwens 
et al. (2006) at $z=6$, Bouwens et al. (2005) at $z=7.5$, and 
Bouwens et al. (2005) at $z=10$.  
The solid circles represent the star formation rate densities achieved 
by integrating
the derived luminosity function down to 0.1 L$^\star_{z=3}$.  
Integrating the luminosity function down to zero luminosity 
(open circles) assuming a faint-end slope of $\alpha=-1.73$
adds an additional factor of 2.3 to the star formation rate density.  
The evolution of the star formation rate density with redshift is 
well fit by a $(1+z)^{-3.3}$ parameterization over 5 $< z <$ 10.  
}
\label{plot:sfrd}
\end{figure}

As \cite{Bou06} discuss, at $z\simeq$6 there is some disagreement in 
the value of the SFRD. The disagreement stems primarily from whether the 
shape of the luminosity function is evolving. Bouwens et al argue for 
a decrease in the characteristic luminosity and an increase in the faint end 
slope prior to z$\simeq$3, so we adopt the redshift-dependent LF parameters 
derived in Bouwens et al. (2006) and Stanway (2005) and integrate accordingly.
We find that the SFRD (integrated down to 0.1 L$^\star_{z=3}$) can be fit 
reasonably well by $\rho  \propto \rm~(1+z)^{-3.3}$ between z=5-10 (Figure \ref{plot:sfrd}).  

\begin{figure}
\plotone{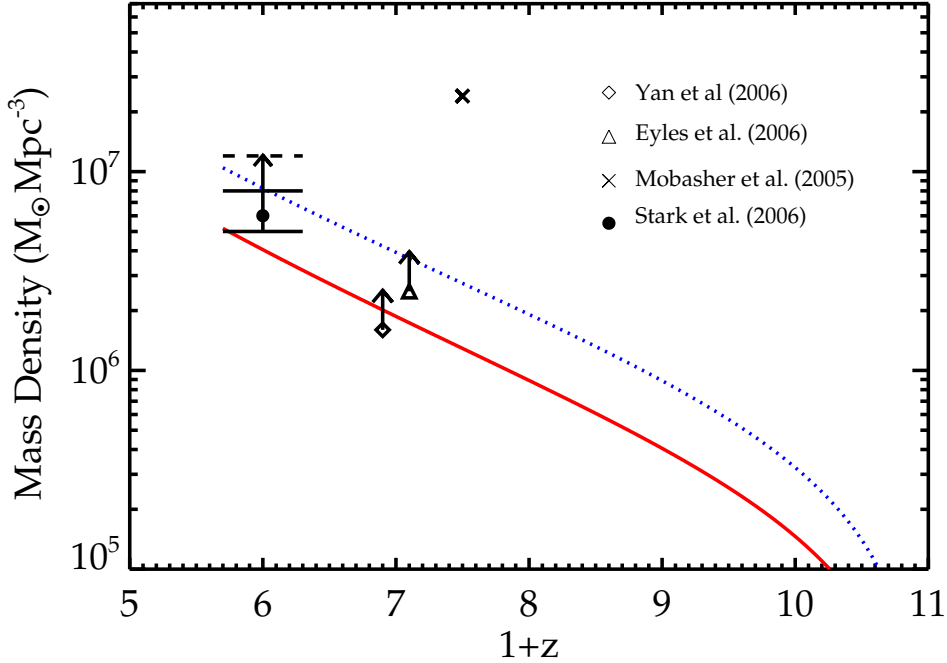}
\caption{Comparison of stellar mass density at $z\simeq $5-10 
derived from spectral energy distributions of galaxies in 
GOODS-S with that derived from integrating the observed
star formation rate density at z$\simeq 5-10$ integrated
down to 0.1 L$^\star$$_{z=3}$ (solid line) and zero luminosity 
(dotted line) assuming a faint-end slope of $\alpha=-$1.73.  The filled 
solid circle with the upward arrow corresponds to the 
stellar mass density derived from spectroscopically-confirmed
galaxies in GOODS-S (1$\times$10$^6$ M$_\odot$) whereas the solid filled 
circle corresponds to the stellar mass density from the photometric 
sample (6$\times$10$^{6}$ M$_\odot$).  We estimate that the stellar 
mass density may be as high as 1$\times$10$^7$ M$_\odot$ Mpc$^{-3}$ 
(solid horizontal line) depending on the contribution from undetected 
sources.  We also include previous 
estimates of the stellar mass density at $z > 5$ from Yan et al. 
(2006) (diamonds), Eyles et al. (2006) (triangle), and 
Mobasher et al. (2005) (cross). The Yan et al. (2006) and Eyles et al. 
(2006) symbols are offset slightly from $z = 6$ for clarity.  The 
large stellar mass density 
at $z\simeq $5 inferred from this study suggests that a significant 
amount of star formation is hidden by dust or has yet to be located, 
perhaps lying at higher redshift or in intrinsically faint systems.}
\label{plot:mass}
\end{figure}

The stellar mass density obtained by integrating this function over 
time between z$\simeq$10 and $z\simeq 5$ is lower than that derived
from the photometric sample of $z\simeq5$ objects in this paper 
(Figure \ref{plot:mass}).  We note that the integral of the star 
formation rate density as a function of redshift overestimates the 
mass density.  This is because we do not account for the mass that is
returned to the interstellar medium in stellar winds and stellar 
deaths.  This can be quantified using the \cite{bc03} software (we 
do this to compute stellar masses, see $\S$5.1), but it is a complicated function 
of the average star formation history and age, which are not well-constrained.
We find that this effect could {\it reduce} the stellar masses inferred from 
integrating the observed star formation rate densities by up to $\simeq$30\% 
(assuming a 1 Gyr instantaneous burst).  Given the uncertainties, we do not 
adjust the curves in Figure 7 by this factor, but we note that this effect further 
enhances the discrepancy between the observed stellar mass density at 
z$\simeq 5$ and that which can be accounted for by previous star formation.
Therefore, the observed stellar mass of the $z\simeq 5$ 
galaxies in GOODS-S either implies a significant amount of dust extinction 
or that not all star formation at z$>$5 has been observed in current surveys.  

To examine the amount of star-formation that may be 
hidden in low-luminosity systems, we integrate the luminosity 
functions to zero luminosity utilizing faint-end slopes of 
$\alpha=-$1.73 (as measured at z$\simeq$6 in \citealt{Bou06}) 
and $\alpha=-$1.9 (as suggested by \citealt{yw04}) and 
integrate the luminosity function to zero luminosity.  

Star formation is unlikely to occur at very low luminosities because of
radiative feedback processes and (after reionization) a photoionized IGM 
which raises the cosmological Jeans mass. We nonetheless take this 
extreme approach to place an upper limit on the amount of unextincted 
star formation.  Assuming no evolution in the shape of the LF between $z\simeq 5$
and $z\simeq 11$, this increases the predicted stellar mass at $z\simeq$ 5 
by an extra factor of 2.3 for faint end slopes of $\alpha=-$1.73;  if we 
instead consider an extreme faint-end slope of $\alpha=-1.9$, the star 
formation rate density increases by a factor of 6.8.  This gives a better 
account of the assembled mass and if correct has interesting consequences 
for higher redshift surveys probing to low luminosities \cite{S07}. 

A significant amount of star formation may also be enshrouded by dust. However, 
recent observations have shown that the rest-frame UV slope of $z\simeq $6 
galaxies is actually somewhat {\it bluer} than that at $z\simeq $3 suggesting
that the mean dust extinction declines between $z\simeq $3-6 
\citep{stan_etal05,Y05,Bou06}.  
Taking the empirically derived fit relating the extinction at 1600~\AA~(A$_{1600}$)
to the UV slope $\beta$, A$_{1600}$=4.43 + 1.99$\beta$ \citep{Me99} yields an overall 
attenuation factor of $\times\simeq$1-1.5 at $z\simeq $6 for two different estimates
of the UV continuum slope at $z\simeq 6$ ($\beta=-2.2$ from \citealt{stan_etal05} and 
$\beta=-2.0$ from \citealt{Bou06}).

Hence, the expected extinction correction to the SFR density at $z\simeq 5-10$ 
could in principle account for the stellar mass contained in the photometric sample 
in this paper if the \cite{Bou06} estimate of the UV continuum slope is correct.  
If there exists either a significant population of quiescent massive galaxies 
or low-mass 
galaxies below the 3$\sigma$ 3.6 $\mu$m flux limit imposed on the data, a significant 
amount of low-luminosity star-forming galaxies would be required 
to assemble the stellar mass.  Future studies will test this hypothesis.

\section{Conclusions}
\label{sec:conclusion}

We have argued that the assembled stellar mass density at high redshift
provides a valuable constraint on the past star formation history and,
with improved precision, may ultimately indicate whether there was sufficient
star formation in the previous $\simeq$500-900 Myr to reionize the
intergalactic medium.

We have demonstrated both the promise and limitations of this method by 
computing the comoving stellar mass density at $z\simeq 5$.  Following
the ideas discussed in \cite{SE2006}, we use the stellar mass density 
to constrain the amount of star formation at earlier times.  

We detail our findings below.

1. We construct a sample of 25 spectroscopically confirmed $z\simeq $5
objects in GOODS-S (14 of which are uncontaminated in the {\em Spitzer} 
data) to place a robust lower limit on the comoving stellar mass density.  
Fitting the SEDs of these objects to templates from \cite{bc03}
populations synthesis models, we infer a total comoving stellar mass density of
$1\times $10$^6$ M$_\odot$ Mpc$^{-3}$.

2. We construct a sample of $z\simeq 5$ galaxies using the 
photometric redshifts of the GOODS-MUSIC catalog. After  
removing likely stellar and low-z contaminants, 196 objects remain in the sample.
Computing the stellar mass from the 72 objects that are uncontaminated by 
nearby sources in the Spitzer data, we estimate a stellar mass density
of 6$\times$10$^6$ M$_\odot$ Mpc$^{-3}$.  Systematic uncertainty in the 
star formation history causes this value to be uncertain at the 30\% level.

3.  The total comoving stellar mass density (6$\times10^6$ M$_\odot$ 
Mpc$^{-3}$) represents a lower limit for several reasons.  First, robust 
stellar mass estimates are only attainable for
reasonably massive galaxies; hence the estimates presented in this paper
do not include the contribution from low-mass systems.  Second, we require
objects to be bright in the rest-frame UV (and hence actively forming stars) 
for selection into our sample.  If there is a large population of quiescent 
galaxies at $z\simeq 5$, the total stellar mass density may be significantly
higher than estimated.  We estimate that the stellar mass density of massive 
galaxies is unlikely to exceed 1$\times$10$^7$ M$_\odot$ Mpc$^{-3}$.

4. The estimated comoving stellar mass density at $z\simeq$ 5 suggests that current 
observations may be missing some star formation 
at z $>$ 5. The missing star formation could, however, be accomodated by 
extincted star formation in LBGs currently seen at $z\simeq$ 6-10 
or in low-luminosity star-forming systems below the detection threshold of current 
observations.  In the latter case, our results have important implications for 
searches for low luminsoity SF systems at high redshift.

\subsection*{ACKNOWLEDGMENTS}

D.P.S. is grateful for the hospitality of the School of 
Physics at the University of Exeter, where most of 
this work was completed.  We thank Peter Capak, Johan Richard, Kevin Bundy, 
Adam Kraus, Elizabeth Stanway, Kuenley Chiu and Richard McMahon for 
enlightening conversations.  We thank 
our anonymous referee for very insightful comments.  A.B. gratefully acknowledges
support from a Philip Leverhulme Prize.  L.E. is supported by 
a PPARC studentship.  This paper is based on
observations made with the NASA/ESA Hubble Space Telescope, obtained
from the Data Archive at the Space Telescope Science Institute, which is
operated by the Association of Universities for Research in Astronomy,
Inc., under NASA contract NAS 5-26555. The {\em HST/ACS} observations are
associated with proposals \#9425\,\&\,9583 (the GOODS public imaging
survey). 
Spitzer
VLT/FORS2
VLT/ISAAC
We are grateful to the GOODS team for making their reduced
images public -- a very useful resource.

\bibliography{journals_apj,mybib}
\begin{landscape}
\begin{deluxetable}{crrcccccccc}
\tablewidth{560pt}
\tabletypesize{\fontsize{7}{10}\selectfont}
\tablecaption{Photometric catalog of $z\simeq 5$ galaxies in
GOODS-S Field\label{tab:vdrop_photom}}
\tablehead{
\colhead{ID} & 
\colhead{RA (J2000)} & 
\colhead{Dec (J2000)} & 
\colhead{v} &  
\colhead{$i'$} & 
\colhead{$z'$} & 
\colhead{$J$} & 
\colhead{$K_s$} & 
\colhead{3.6 $\mu$m} & 
\colhead{4.5 $\mu$m}  &
\colhead{z$_{phot}$}
}
\startdata
  44\_2919 & 03 32 9.054 & -27 43 51.85 &  27.58$\pm$0.16 &  26.02$\pm$0.08 &  25.65$\pm$0.07 &  25.37$\pm$0.26 &  25.50$\pm$0.44 &  24.01$\pm$0.09 &  24.29$\pm$0.14 & 4.590 \\
  42\_3601 & 03 32 10.64 & -27 50 29.15 &  26.83$\pm$0.08 &  25.69$\pm$0.06 &  25.60$\pm$0.07 &  26.04$\pm$0.42 &  25.37$\pm$0.37 &  24.93$\pm$0.20 &  25.41$\pm$0.43 & 4.480 \\
  33\_4001 & 03 32 11.44 & -27 47 38.63 &  27.88$\pm$0.18 &  25.95$\pm$0.06 &  25.77$\pm$0.07 &  25.43$\pm$0.31 &  25.46$\pm$0.71 &  24.00$\pm$0.09 &  24.31$\pm$0.17 & 4.720 \\
  33\_4496 & 03 32 12.42 & -27 47 2.483 &  27.83$\pm$0.20 &  26.35$\pm$0.10 &  26.21$\pm$0.12 &  26.35$\pm$0.59 &  26.03$\pm$0.98 &      $>$25.8    &      $>$25.4    & 4.480 \\
  33\_4687 & 03 32 12.78 & -27 48 2.599 &  27.47$\pm$0.11 &  26.15$\pm$0.07 &  25.96$\pm$0.08 &  25.48$\pm$0.26 &  25.58$\pm$0.62 &  24.29$\pm$0.10 &  24.95$\pm$0.23 & 4.590 \\
  34\_4915 & 03 32 13.25 & -27 43 8.289 &  27.90$\pm$0.22 &  26.32$\pm$0.10 &  26.16$\pm$0.11 &  25.93$\pm$0.41 &  25.85$\pm$0.68 &  25.59$\pm$0.41 &      $>$25.4    & 4.680 \\
  35\_5207 & 03 32 13.88 & -27 41 48.54 &  29.35$\pm$0.83 &  27.27$\pm$0.23 &  26.49$\pm$0.15 &  29.30$\pm$9.44 &      $>$25.3    &  25.88$\pm$0.61 &      $>$25.4    & 5.340 \\
  33\_5533 & 03 32 14.49 & -27 49 32.69 &  26.67$\pm$0.07 &  25.69$\pm$0.06 &  25.53$\pm$0.06 &  26.21$\pm$0.56 &  25.53$\pm$0.48 &  25.07$\pm$0.33 &  25.85$\pm$0.72 & 4.460 \\
  33\_5986 & 03 32 15.35 & -27 49 36.08 &  27.78$\pm$0.19 &  26.15$\pm$0.09 &  26.35$\pm$0.13 &  25.79$\pm$0.38 &  25.40$\pm$0.41 &  25.27$\pm$0.58 &  26.18$\pm$0.44 & 4.610 \\
  33\_6438 & 03 32 16.17 & -27 46 41.59 &  28.74$\pm$0.47 &  26.25$\pm$0.10 &  26.00$\pm$0.10 &  26.63$\pm$0.75 &     $>$25.3     &  25.16$\pm$0.22 &  26.04$\pm$0.64 & 5.080 \\
  33\_6440 & 03 32 16.17 & -27 48 19.42 &  27.83$\pm$0.20 &  26.39$\pm$0.11 &  26.25$\pm$0.13 &  25.71$\pm$0.33 &  25.00$\pm$0.38 &  24.88$\pm$0.15 &  25.28$\pm$0.31 & 4.700 \\
  33\_6519 & 03 32 16.34 & -27 48 31.99 &  27.87$\pm$0.21 &  26.06$\pm$0.08 &  26.13$\pm$0.11 &      $>$26.0    &     $>$25.3     &  25.79$\pm$0.31 &  26.19$\pm$0.68 & 4.490 \\
  33\_6575 & 03 32 16.45 & -27 46 39.24 &  29.45$\pm$0.89 &  26.49$\pm$0.12 &  26.11$\pm$0.11 &  25.90$\pm$0.39 &  25.08$\pm$0.41 &  24.22$\pm$0.09 &  25.00$\pm$0.25 & 5.210 \\
  32\_6854 & 03 32 16.98 & -27 51 23.17 &  27.41$\pm$0.14 &  25.62$\pm$0.05 &  25.70$\pm$0.08 &  25.97$\pm$0.44 &  24.48$\pm$0.19 &  23.97$\pm$0.07 &  24.44$\pm$0.16 & 4.550 \\
  35\_6867 & 03 32 17.00 & -27 41 13.71 &  26.89$\pm$0.08 &  25.38$\pm$0.04 &  25.13$\pm$0.04 &  24.92$\pm$0.23 &  24.37$\pm$0.24 &  23.43$\pm$0.04 &  23.82$\pm$0.08 & 4.590 \\
  32\_8020 & 03 32 18.91 & -27 53 2.746 &  27.77$\pm$0.19 &  25.13$\pm$0.03 &  24.49$\pm$0.03 &  24.74$\pm$0.13 &  24.06$\pm$0.13 &  22.73$\pm$0.02 &  22.74$\pm$0.03 & 5.550 \\
  31\_8593 & 03 32 19.96 & -27 54 58.98 &  28.64$\pm$0.56 &  26.84$\pm$0.21 &  25.83$\pm$0.11 &  25.50$\pm$0.34 &  25.59$\pm$0.45 &  24.66$\pm$0.11 &  25.23$\pm$0.26 & 5.320 \\
  31\_9014 & 03 32 20.70 & -27 55 36.14 &  26.70$\pm$0.10 &  25.67$\pm$0.08 &  25.72$\pm$0.10 &  25.99$\pm$0.54 &  24.99$\pm$0.26 &  24.08$\pm$0.07 &  24.32$\pm$0.12 & 4.520 \\
  33\_9184 & 03 32 21.01 & -27 49 59.16 &  27.23$\pm$0.12 &  26.03$\pm$0.08 &  26.16$\pm$0.11 &  26.67$\pm$0.63 &    $>$25.3      &     $>$25.8     &     $>$25.4     & 4.580 \\
  33\_9338 & 03 32 21.28 & -27 49 59.67 &  28.54$\pm$0.38 &  26.84$\pm$0.16 &  26.14$\pm$0.11 &  25.85$\pm$0.30 &  26.64$\pm$0.92 &  25.76$\pm$0.26 &  25.35$\pm$0.26 & 5.500 \\
  33\_9677 & 03 32 21.82 & -27 50 3.346 &  28.74$\pm$0.46 &  26.49$\pm$0.12 &  26.48$\pm$0.15 &      $>$26.0    &  25.48$\pm$0.31 &  25.56$\pm$0.28 &     $>$25.4     & 4.800 \\
  34\_9738 & 03 32 21.93 & -27 45 33.07 &  28.22$\pm$0.29 &  26.20$\pm$0.09 &  25.82$\pm$0.09 &  26.20$\pm$0.51 &  24.90$\pm$0.23 &  24.31$\pm$0.09 &  24.72$\pm$0.18 & 4.800 \\
  33\_9812 & 03 32 22.02 & -27 46 42.89 &  26.76$\pm$0.08 &  25.41$\pm$0.05 &  25.26$\pm$0.05 &  25.94$\pm$0.48 &  24.84$\pm$0.25 &  23.84$\pm$0.06 &  24.05$\pm$0.10 & 4.510 \\
  34\_9822 & 03 32 22.03 & -27 45 29.31 &  27.38$\pm$0.14 &  26.41$\pm$0.11 &  26.28$\pm$0.13 &  26.00$\pm$0.42 &     $>$25.3     &  25.30$\pm$0.22 &  25.42$\pm$0.33 & 4.570 \\
 33\_10064 & 03 32 22.44 & -27 47 46.17 &  28.58$\pm$0.40 &  26.64$\pm$0.14 &  26.34$\pm$0.13 &  26.46$\pm$0.77 &  25.57$\pm$0.47 &  24.86$\pm$0.13 &  25.22$\pm$0.27 & 5.020 \\
 32\_10232 & 03 32 22.71 & -27 51 54.40 &  27.90$\pm$0.25 &  26.14$\pm$0.10 &  25.68$\pm$0.08 &  25.58$\pm$0.28 &  25.03$\pm$0.24 &  24.27$\pm$0.08 &  24.82$\pm$0.16 & 5.050 \\
 33\_10340 & 03 32 22.88 & -27 47 27.56 &  26.64$\pm$0.07 &  24.94$\pm$0.04 &  24.84$\pm$0.04 &  24.55$\pm$0.13 &  24.59$\pm$0.16 &  23.75$\pm$0.05 &  24.01$\pm$0.10 & 4.440 \\
 31\_10974 & 03 32 24.00 & -27 54 59.79 &  27.51$\pm$0.16 &  25.46$\pm$0.05 &  24.73$\pm$0.03 &  24.56$\pm$0.16 &  24.89$\pm$0.25 &  25.39$\pm$0.41 &  25.88$\pm$0.66 & 5.380 \\
 32\_11635 & 03 32 25.02 & -27 50 24.49 &  29.10$\pm$0.56 &  27.13$\pm$0.18 &  26.05$\pm$0.09 &  25.61$\pm$0.24 &  25.42$\pm$0.30 &  24.68$\pm$0.13 &  24.67$\pm$0.17 & 5.430 \\
 33\_13701 & 03 32 27.94 & -27 46 18.57 &  26.37$\pm$0.06 &  25.18$\pm$0.04 &  25.22$\pm$0.05 &  25.09$\pm$0.21 &  24.44$\pm$0.15 &  24.02$\pm$0.10 &  24.13$\pm$0.13 & 4.480 \\
 34\_14195 & 03 32 28.70 & -27 42 28.95 &  28.11$\pm$0.21 &  26.23$\pm$0.08 &  26.03$\pm$0.08 &  26.82$\pm$0.84 &      $>$25.3    &  25.67$\pm$0.34 &      $>$25.4    & 4.840 \\
 23\_15316 & 03 32 30.28 & -27 49 22.01 &  28.04$\pm$0.25 &  26.47$\pm$0.12 &  25.88$\pm$0.09 &  26.30$\pm$0.47 &     $>$25.3     &    $>$25.8      &      $>$25.4    & 5.200 \\
 22\_15851 & 03 32 31.07 & -27 51 17.85 &  28.91$\pm$0.54 &  26.17$\pm$0.09 &  26.04$\pm$0.10 &  26.06$\pm$0.40 &  24.89$\pm$0.22 &  24.62$\pm$0.17 &  24.93$\pm$0.25 & 4.820 \\
 23\_16055 & 03 32 31.37 & -27 48 13.81 &  28.08$\pm$0.26 &  26.20$\pm$0.10 &  25.89$\pm$0.09 &  25.09$\pm$0.20 &  25.66$\pm$0.39 &  24.94$\pm$0.14 &  25.42$\pm$0.32 & 4.990 \\
 22\_17535 & 03 32 33.69 & -27 53 21.62 &  29.06$\pm$0.61 &  27.28$\pm$0.24 &  26.39$\pm$0.14 &      $>$26.0    &     $>$25.3     &  26.28$\pm$0.60 &     $>$25.4     & 5.360 \\
 23\_17728 & 03 32 33.98 & -27 48 2.043 &  27.61$\pm$0.17 &  26.10$\pm$0.09 &  25.90$\pm$0.09 &  25.74$\pm$0.29 &  24.52$\pm$0.13 &  24.08$\pm$0.07 &  24.24$\pm$0.11 & 4.470 \\
 23\_18716 & 03 32 35.45 & -27 49 35.20 &  29.41$\pm$0.60 &  26.63$\pm$0.10 &  26.35$\pm$0.09 &  25.71$\pm$0.28 &  25.94$\pm$0.49 &  25.12$\pm$0.22 &  25.91$\pm$0.56 & 4.930 \\
 22\_19011 & 03 32 35.89 & -27 52 44.02 &  27.78$\pm$0.19 &  26.51$\pm$0.12 &  26.27$\pm$0.12 &  26.92$\pm$0.85 &     $>$25.3     &  25.70$\pm$0.34 &    $>$25.4      & 4.740 \\
 24\_19118 & 03 32 36.08 & -27 44 3.942 &  27.28$\pm$0.12 &  26.07$\pm$0.08 &  25.75$\pm$0.08 &  24.80$\pm$0.12 &  25.22$\pm$0.26 &  24.38$\pm$0.13 &  25.01$\pm$0.27 & 4.550 \\
 23\_19268 & 03 32 36.30 & -27 49 52.79 &  27.44$\pm$0.14 &  26.03$\pm$0.08 &  25.86$\pm$0.09 &  25.36$\pm$0.21 &  24.70$\pm$0.16 &  23.96$\pm$0.09 &  24.56$\pm$0.18 & 4.510 \\
 24\_19435 & 03 32 36.49 & -27 43 53.46 &  28.52$\pm$0.37 &  26.93$\pm$0.17 &  26.43$\pm$0.14 &  25.83$\pm$0.31 &  25.93$\pm$0.55 &  25.01$\pm$0.19 &  25.59$\pm$0.42 & 4.620 \\
 25\_19912 & 03 32 37.25 & -27 42 2.570 &  28.47$\pm$0.35 &  26.42$\pm$0.11 &  26.00$\pm$0.10 &  25.81$\pm$0.32 &    $>$25.3      &  25.04$\pm$0.18 &  25.53$\pm$0.41 & 5.180 \\
 22\_20159 & 03 32 37.62 & -27 50 22.38 &     $>$29.5     &  27.18$\pm$0.22 &  26.24$\pm$0.12 &  25.37$\pm$0.20 &  25.54$\pm$0.36 &  24.64$\pm$0.14 &  24.66$\pm$0.17 & 5.510 \\
 22\_20304 & 03 32 37.86 & -27 52 59.10 &  27.43$\pm$0.15 &  26.47$\pm$0.12 &  26.25$\pm$0.13 &     $>$26.0     &    $>$25.3      &  25.68$\pm$0.45 &  25.92$\pm$0.57 & 4.520 \\
 23\_20360 & 03 32 37.95 & -27 47 11.05 &  27.39$\pm$0.13 &  25.80$\pm$0.07 &  26.21$\pm$0.12 &     $>$26.0     &    $>$25.3      &  24.93$\pm$0.19 &     $>$25.4     & 4.690 \\
 22\_21669 & 03 32 40.08 & -27 50 49.60 &  27.72$\pm$0.18 &  26.28$\pm$0.10 &  26.05$\pm$0.10 &  25.74$\pm$0.30 &  25.56$\pm$0.40 &  25.45$\pm$0.29 &  26.17$\pm$0.64 & 4.520 \\
 24\_22091 & 03 32 40.85 & -27 45 46.25 &  28.03$\pm$0.18 &  26.14$\pm$0.07 &  25.33$\pm$0.04 &  25.12$\pm$0.14 &  25.22$\pm$0.32 &  24.76$\pm$0.12 &  24.45$\pm$0.14 & 5.400 \\
 23\_22354 & 03 32 41.34 & -27 48 43.13 &  28.02$\pm$0.24 &  26.31$\pm$0.11 &  26.14$\pm$0.11 &     $>$26.0     &  26.02$\pm$0.62 &  24.72$\pm$0.13 &  25.46$\pm$0.33 & 4.530 \\
 25\_22925 & 03 32 42.36 & -27 41 14.87 &  27.38$\pm$0.21 &  25.94$\pm$0.13 &  25.41$\pm$0.09 &  26.10$\pm$0.62 &  24.65$\pm$0.36 &  24.12$\pm$0.09 &  24.77$\pm$0.21 & 5.010 \\
 24\_23215 & 03 32 42.95 & -27 43 39.65 &  29.16$\pm$0.67 &  26.68$\pm$0.14 &  26.09$\pm$0.11 &     $>$26.0     &    $>$25.3      &  25.94$\pm$0.59 &  25.84$\pm$0.63 & 5.200 \\
 24\_23395 & 03 32 43.30 & -27 43 10.59 &  28.73$\pm$0.45 &  26.14$\pm$0.09 &  26.25$\pm$0.12 &  25.60$\pm$0.32 &    $>$25.3      &  24.74$\pm$0.16 &  25.10$\pm$0.30 & 4.780 \\
 23\_23515 & 03 32 43.53 & -27 49 19.21 &  27.89$\pm$0.21 &  25.80$\pm$0.07 &  25.53$\pm$0.07 &  26.05$\pm$0.34 &  25.84$\pm$0.50 &  24.89$\pm$0.15 &  25.92$\pm$0.49 & 5.090 \\
 22\_25323 & 03 32 47.58 & -27 52 28.18 &  28.49$\pm$0.34 &  26.43$\pm$0.11 &  26.24$\pm$0.11 &  25.43$\pm$0.26 &    $>$25.3      &  25.43$\pm$0.29 &     $>$25.4     & 4.760 \\
 13\_25544 & 03 32 48.14 & -27 48 17.69 &  27.95$\pm$0.23 &  26.04$\pm$0.09 &  25.41$\pm$0.06 &  25.64$\pm$0.28 &  24.89$\pm$0.25 &  24.65$\pm$0.12 &  24.97$\pm$0.24 & 5.170 \\
 14\_25620 & 03 32 48.33 & -27 45 38.90 &  27.86$\pm$0.21 &  26.24$\pm$0.11 &  26.37$\pm$0.14 &  26.42$\pm$0.49 &  25.75$\pm$0.52 &     $>$25.8     &    $>$25.4      & 4.610 \\
 12\_25696 & 03 32 48.53 & -27 54 25.67 &  27.44$\pm$0.12 &  26.28$\pm$0.08 &  26.24$\pm$0.10 &  26.49$\pm$0.77 &    $>$25.3      &  24.89$\pm$0.20 &  25.63$\pm$0.45 & 4.530 \\
 12\_25851 & 03 32 48.89 & -27 52 43.17 &  27.75$\pm$0.18 &  26.49$\pm$0.12 &  26.39$\pm$0.14 &     $>$26.0     &    $>$25.3      &     $>$25.8     &    $>$25.4      & 4.540 \\
 12\_25952 & 03 32 49.15 & -27 50 22.52 &  28.24$\pm$0.29 &  26.04$\pm$0.08 &  25.30$\pm$0.05 &  25.27$\pm$0.19 &  25.52$\pm$0.39 &  25.20$\pm$0.18 &  25.54$\pm$0.33 & 5.360 \\
 12\_26198 & 03 32 49.81 & -27 50 22.75 &  28.23$\pm$0.28 &  26.42$\pm$0.10 &  26.12$\pm$0.10 &  25.58$\pm$0.27 &  25.35$\pm$0.35 &  25.94$\pm$0.41 &  26.22$\pm$0.65 & 5.110 \\
 12\_26409 & 03 32 50.44 & -27 50 39.64 &     $>$29.5     &  27.01$\pm$0.20 &  26.38$\pm$0.14 &    $>$26.0      &      $>$25.3    &  26.30$\pm$0.57 &  26.70$\pm$0.99 & 5.460 \\
 13\_26480 & 03 32 50.63 & -27 49 34.79 &     $>$29.5     &  26.83$\pm$0.16 &  26.36$\pm$0.14 &    $>$26.0      &      $>$25.3    &     $>$25.8     &    $>$25.4      & 5.220 \\
 13\_26492 & 03 32 50.65 & -27 47 15.18 &  26.80$\pm$0.09 &  25.84$\pm$0.08 &  25.64$\pm$0.07 &  26.26$\pm$0.49 &      $>$25.3    &  25.42$\pm$0.28 &    $>$25.4      & 4.430 \\
 12\_26985 & 03 32 51.94 & -27 52 8.494 &  27.46$\pm$0.15 &  26.04$\pm$0.09 &  25.95$\pm$0.09 &  26.19$\pm$0.58 &     $>$25.3     &    $>$25.8      &    $>$25.4      & 4.580 \\
 12\_27749 & 03 32 54.05 & -27 51 12.02 &  29.12$\pm$0.68 &  27.09$\pm$0.20 &  26.32$\pm$0.13 &    $>$26.0      &     $>$25.3     &  25.82$\pm$0.85 &    $>$25.4      & 5.500 \\
 12\_28370 & 03 32 56.22 & -27 51 51.29 &  27.81$\pm$0.20 &  26.33$\pm$0.10 &  26.17$\pm$0.11 &    $>$26.0      &     $>$25.3     &  25.01$\pm$0.21 &    $>$25.4      & 4.570 \\
 12\_28389 & 03 32 56.29 & -27 53 31.53 &  27.33$\pm$0.12 &  25.66$\pm$0.05 &  25.60$\pm$0.07 &  24.62$\pm$0.17 &  24.83$\pm$0.27 &  24.61$\pm$0.15 &  25.03$\pm$0.29 & 4.690 \\
 12\_28728 & 03 32 57.68 & -27 53 19.67 &  28.05$\pm$0.25 &  26.09$\pm$0.08 &  26.08$\pm$0.10 &  25.99$\pm$0.59 &    $>$25.3      &  24.99$\pm$0.21 &  25.26$\pm$0.32 & 4.930 \\
 12\_28859 & 03 32 58.38 & -27 53 39.59 &  26.41$\pm$0.05 &  25.54$\pm$0.04 &  25.62$\pm$0.06 &    $>$26.0      &  26.65$\pm$0.88 &  24.98$\pm$0.30 &  24.92$\pm$0.28 & 4.420 \\
 12\_28917 & 03 32 58.66 & -27 52 43.69 &  28.33$\pm$0.32 &  26.11$\pm$0.09 &  25.81$\pm$0.08 &    $>$26.0      &  24.98$\pm$0.33 &  25.05$\pm$0.20 &  25.94$\pm$0.64 & 4.840 \\
 12\_28990 & 03 32 59.01 & -27 53 32.22 &  27.47$\pm$0.13 &  25.55$\pm$0.05 &  25.20$\pm$0.04 &  25.13$\pm$0.21 &  24.42$\pm$0.17 &  23.39$\pm$0.06 &  24.07$\pm$0.13 & 4.860 \\
 12\_29097 & 03 32 59.72 & -27 52 2.582 &  29.20$\pm$0.73 &  26.31$\pm$0.10 &  25.67$\pm$0.07 &    $>$26.0      &    $>$25.3      &  24.05$\pm$0.11 &  24.66$\pm$0.25 & 5.170 \\
 12\_29119 & 03 32 59.89 & -27 52 56.42 &  28.53$\pm$0.40 &  26.70$\pm$0.15 &  26.29$\pm$0.13 &    $>$26.0      &  24.91$\pm$0.15 &  23.88$\pm$0.08 &  24.12$\pm$0.14 & 4.890 \\
\enddata
\end{deluxetable}
\end{landscape}

\begin{deluxetable}{crrcccccc}
\tabletypesize{\fontsize{7}{10}\selectfont}
\tablecaption{Spectroscopically-confirmed $z\simeq$5 galaxies in
GOODS-S Field\label{tab:spec_sample}}
\tablehead{
\colhead{ID} & 
\colhead{RA (J2000)} & 
\colhead{Dec (J2000)} & 
\colhead{Redshift} &
\colhead{$v-i'$} &
\colhead{$i'-z'$} &
\colhead{Spitzer Confusion Class} &
\colhead{Redshift Flag} &
\colhead{Included in SED fitting}
}
\startdata
  44\_1543 & 03 32 5.258 & -27 43 0.406  & 4.81  &  1.85  & -0.09  & 4 &  A & N  \\
  35\_4142 & 03 32 11.71 & -27 41 49.59  & 4.91  &  1.79  &  0.17  & 3 &  C & Y  \\
  35\_4244 & 03 32 11.92 & -27 41 57.09  & 5.57  &  1.39  &  1.05  & 3/4 &  B & N  \\
  35\_6626 & 03 32 16.55 & -27 41 3.203  & 5.25  &  2.07  &  0.92  & 3 &  C & Y  \\
  35\_6867 & 03 32 17.00 & -27 41 13.71  & 4.41  &  1.54  &  0.11  & 1 &  B & N  \\
  33\_7471 & 03 32 17.95 & -27 48 17.01  & 5.40  &  1.82  &  1.16  & 4 &  C & N  \\
  32\_8020 & 03 32 18.91 & -27 53 2.746  & 5.55  &  2.60  &  0.57  & 1 &  A & Y  \\
  35\_9350 & 03 32 21.30 & -27 40 51.20  & 5.29  &  1.84  &  0.71  & 1 &  A & Y  \\
  34\_9738 & 03 32 21.93 & -27 45 33.07  & 4.78  &  2.00  &  0.28  & 3 &  C & Y  \\
 32\_10232 & 03 32 22.71 & -27 51 54.40  & 4.90  &  2.05  &  0.42  & 1 &  C & Y  \\
 33\_10340 & 03 32 22.88 & -27 47 27.56  & 4.44  &  1.60  &  0.00  & 1 &  B & Y  \\
  no\_ACS\_01 & 03 32 22.89 & -27 45 20.99  & 5.12  &  \nodata  & \nodata & \nodata & C & N  \\
 33\_10388 & 03 32 22.97 & -27 46 29.09  & 4.50  &  1.65  &  0.03  & 3/4 & C & N  \\
 34\_11820 & 03 32 25.31 & -27 45 30.85  & 4.99  &  3.55  &  0.45  & 2   & B & Y  \\
 35\_14097 & 03 32 28.56 & -27 40 55.71  & 4.59  &  1.67  &  0.37  & 3   & B & Y  \\
 35\_14303 & 03 32 28.84 & -27 41 32.70  & 4.80  &  1.79  &  0.02  & 4   & B & N  \\
  no\_ACS\_02 & 03 32 28.93 & -27 41 28.19  & 4.88  &  \nodata  & \nodata & \nodata & B & N  \\
 31\_14602 & 03 32 29.29 & -27 56 19.46  & 4.76  &  1.67  &  0.10  & 3   & B & Y  \\
 22\_15184 & 03 32 30.09 & -27 50 57.72  & 5.08  &  0.39  & -0.05  & \nodata & B & N  \\
 24\_18073 & 03 32 34.48 & -27 44 3.008  & 4.94  &  1.46  & -0.12  & 3/4  & C & N  \\
 22\_20159 & 03 32 37.62 & -27 50 22.38  & 5.51  &  1.89  &  1.28  & 3/4  & A & N  \\
 22\_21502 & 03 32 39.81 & -27 52 58.09  & 5.54  &  1.63  &  1.03  &  4   & C & N  \\
 24\_21686 & 03 32 40.11 & -27 45 35.49  & 4.77  &  1.62  &  0.16  & 3/4 & B & N  \\
 21\_23040 & 03 32 42.62 & -27 54 28.95  & 4.40  &  1.85  &  0.41  & 3   & C & Y  \\
 23\_23051 & 03 32 42.65 & -27 49 38.99  & 4.84  &  2.23  &  0.06  & 3   & C & Y  \\
  no\_ACS\_03 & 03 32 43.15 & -27 50 34.80  & 4.83  &  \nodata  & \nodata & \nodata & C & N  \\
 23\_24305 & 03 32 45.23 & -27 49 9.829  & 5.58  &  1.61  &  1.22 & 3/4 & B &  N  \\
 21\_24396 & 03 32 45.42 & -27 54 38.52  & 5.37  &  2.64  &  0.69 & 3   & A &  Y  \\
 22\_25323 & 03 32 47.58 & -27 52 28.18  & 4.75  &  2.01  &  0.11 & 3   & C &  Y  \\
 12\_28085 & 03 32 55.08 & -27 54 14.48  & 4.71  &  1.88  &  0.13 & 3/4 & A &  N  \\
\enddata
\tablecomments{All magnitudes are in AB system.  no\_ACS\_01, no\_ACS\_02, and 
no\_ACS\_03, were not detected with {\em ACS}.  The {\em Spitzer} confusion 
classes have the following meanings: 1 = isolated and detected; 2 = isolated but undetected;
3 = confused, but Galfit may help ; 4 = hopelessly confused.  Those sources with Spitzer
confusion class '3/4' were deemed hopelessly confused after attempting (and 
failing) to subtract nearby sources with Galfit.}
\end{deluxetable}

\clearpage

\begin{deluxetable}{lcccccccc}
\tabletypesize{\fontsize{7}{10}\selectfont}
\tablecaption{Photometric Properties 
of z$\simeq$5  spectroscopically-confirmed 
galaxies. \label{tab:spec_photom}}
\tablehead{
\colhead{ID} &  
\colhead{z$_{spec}$} &  
\colhead{v} &  
\colhead{$i'$} & 
\colhead{$z'$} & 
\colhead{$J$} & 
\colhead{$K_s$} & 
\colhead{3.6 $\mu$m} & 
\colhead{4.5 $\mu$m}}
\startdata    
  35\_4142 & 4.912 &  27.22$\pm$0.11 &  25.51$\pm$0.05 &  25.26$\pm$0.05 &  25.19$\pm$0.22 &  25.09$\pm$0.35 &  25.11$\pm$0.27 &  25.65$\pm$0.54   \\
  35\_6626 & 5.250 &  29.07$\pm$0.62 &  27.18$\pm$0.21 &  26.35$\pm$0.13 &  \nodata        &  \nodata        &  26.19$\pm$0.52 &
$>$25.4  
       \\
  35\_6867 & 4.416 &  26.89$\pm$0.08 &  25.38$\pm$0.04 &  25.13$\pm$0.04 &  24.92$\pm$0.23 &  24.37$\pm$0.24 &  23.43$\pm$0.04 &  23.82$\pm$0.08  
       \\
  32\_8020 & 5.554 &  27.77$\pm$0.19 &  25.13$\pm$0.03 &  24.49$\pm$0.03 &  24.74$\pm$0.13 &  24.06$\pm$0.13 &  22.73$\pm$0.02 &  22.74$\pm$0.03  
       \\
  35\_9350 & 5.283 &  28.07$\pm$0.32 &  26.04$\pm$0.10 &  25.41$\pm$0.08 &  \nodata        &  \nodata        &  \nodata        &  25.52$\pm$0.60  
       \\
  34\_9738 & 4.788 &  28.22$\pm$0.29 &  26.20$\pm$0.09 &  25.82$\pm$0.09 &  26.20$\pm$0.51 &  24.90$\pm$0.23 &  24.31$\pm$0.09 &  24.72$\pm$0.18         \\
 32\_10232 & 4.900 &  27.90$\pm$0.25 &  26.14$\pm$0.10 &  25.68$\pm$0.08 &  25.58$\pm$0.28 &  25.03$\pm$0.24 &  24.27$\pm$0.08 &  24.82$\pm$0.16  
       \\
 33\_10340 & 4.440 &  26.64$\pm$0.07 &  24.94$\pm$0.04 &  24.84$\pm$0.04 &  24.55$\pm$0.13 &  24.59$\pm$0.16 &  23.75$\pm$0.05 &  24.01$\pm$0.10  
       \\
 34\_11820 & 4.992 &  28.78$\pm$0.45 &  26.95$\pm$0.16 &  26.66$\pm$0.16 &  $>$26.0        &  25.63$\pm$0.46 &  $>$25.8        &  $>$25.4  
       \\
 35\_14097 & 4.597 &  27.72$\pm$0.17 &  25.92$\pm$0.07 &  25.87$\pm$0.08 &  \nodata        &  \nodata        &  25.19$\pm$0.30 &  25.16$\pm$0.38         \\
 31\_14602 & 4.760 &  26.76$\pm$0.09 &  25.04$\pm$0.04 &  24.88$\pm$0.04 &  \nodata        &  \nodata        &  22.64$\pm$0.02 &  22.55$\pm$0.02  
       \\
 21\_23040 & 4.400 &  28.17$\pm$0.28 &  26.12$\pm$0.08 &  25.79$\pm$0.08 &  25.30$\pm$0.26 &  25.49$\pm$0.43 &  24.02$\pm$0.06 &  24.57$\pm$0.14  
       \\
 23\_23051 & 4.840 &  28.31$\pm$0.31 &  26.04$\pm$0.08 &  25.86$\pm$0.09 &  26.19$\pm$0.38 &  25.87$\pm$0.50 &  24.93$\pm$0.14 &  25.12$\pm$0.23  
       \\
 21\_24396 & 5.370 &  28.94$\pm$0.55 &  26.09$\pm$0.08 &  25.30$\pm$0.05 &  24.99$\pm$0.22 &  25.11$\pm$0.33 &  24.80$\pm$0.16 &  24.21$\pm$0.11  
       \\
 22\_25323 & 4.758 &  28.49$\pm$0.34 &  26.43$\pm$0.11 &  26.24$\pm$0.11 &  25.43$\pm$0.26 &  $>$25.3        &  25.43$\pm$0.29 &  $>$25.4  
       \\
\enddata
\tablecomments{The VLT mosaic does not cover the entire GOODS field.  Those sources that are 
located off the edge of the VLT images denoted by an ellipsis.  } 
\end{deluxetable}

\begin{deluxetable}{lrrrrr}
\tablecaption{Modeling Results\label{tab:mod}}
\tablehead{
\colhead{Object} &  
\colhead{Star Formation History} & 
\colhead{log$\rm~M_{stellar} (~M_\odot$)} & 
\colhead{Age (Myr)} & 
\colhead{E(B-V)} & 
\colhead{$\chi^2$}}
\startdata
 35\_4142 &   csf &  9.34 &  161 & 0.00 & 2.11 \\
 35\_6626 &    70 &  9.32 &  143 & 0.00 & 0.26 \\
 35\_6867 &   100 & 10.37 &  360 & 0.01 & 1.77 \\
 32\_8020 &   300 & 11.16 &  905 & 0.00 & 2.90 \\
 35\_9350 &   csf &  9.33 &  255 & 0.00 & 0.57 \\
 34\_9738 &   100 & 10.13 &  360 & 0.00 & 1.37 \\
32\_10232 &    70 & 10.06 &  255 & 0.01 & 2.22 \\
33\_10340 &   100 & 11.28 &   18 & 0.24 & 1.82 \\
35\_14097 &   200 &  9.93 &  255 & 0.05 & 0.04 \\
31\_14602 &   300 & 11.10 & 1015 & 0.00 & 1.79 \\
21\_23040 & burst &  8.43 &    1 & 0.53 & 2.24 \\
23\_23051 &   100 &  9.86 &  286 & 0.00 & 0.47 \\
21\_24396 & burst &  8.40 &    9 & 0.17 & 0.21 \\
22\_25323 & burst &  8.43 &    3 & 0.32 & 0.61 \\
\enddata
\tablecomments{In the star-formation history column,
'csf' corresponds to constant star formation, while 
the numbers (e.g 70, 100) correspond to the exponential decay 
constant (in Myr) for an exponentially-declining 
star formation history.
}
\end{deluxetable}

\clearpage

\clearpage

\clearpage

\clearpage

\clearpage

\end{document}